\documentclass[11pt,reqno]{amsart}

\usepackage{graphics,amsmath,amssymb,epsfig,stmaryrd,color,amsaddr}
\usepackage{amsfonts,array}
\usepackage[hmargin=2.5cm, vmargin=2.54cm]{geometry}
\usepackage{graphicx}
\graphicspath{ {./figures/} {./images/} }
\usepackage{amsthm}
\usepackage{mathabx}
\usepackage{flafter}
\usepackage[svgnames]{xcolor}
\usepackage{bbm}
\usepackage{booktabs}
\usepackage{float}
\newcommand{\indep}{\rotatebox[origin=c]{90}{$\models$}}
\usepackage{natbib}
\usepackage{ulem}
\usepackage{enumerate}
\usepackage{threeparttable}
\usepackage{threeparttablex}
\usepackage{multirow}
\usepackage{subcaption}
\usepackage{tabularx}
\usepackage[font=bf, justification=centering, width=\textwidth]{caption}
\usepackage[nodisplayskipstretch]{setspace}

\newcolumntype{H}{>{\setbox0=\hbox\bgroup}c<{\egroup}@{}}
\newcolumntype{P}[1]{>{\centering\arraybackslash}p{#1}}
\newcolumntype{Y}{>{\centering\arraybackslash}X}

\makeatletter
\renewcommand\@endtheorem{\vvv@endmarker\endtrivlist\@endpefalse}
\newcommand\vvv@endmarker{%
  {\nobreak\hfil\penalty50
  \hskip2em\vadjust{}\nobreak\hfil\openbox
  \parfillskip=0pt \finalhyphendemerits=0 \par
  \penalty 10000 \parskip=0pt\noindent}\ignorespaces}
\makeatother

\newtheorem{theorem}{Theorem}
\theoremstyle{plain}

\newtheorem*{assumption*}{Assumption}
\newtheorem{assumption}{Assumption}

\newtheorem{corollary}{Corollary}

\newtheorem{lemma}{Lemma}
\newtheorem{notation}{Notation}

\newtheorem{proposition}{Proposition}

\numberwithin{equation}{section}

\usepackage{hyperref}
\definecolor{darkblue}{rgb}{0.0,0.0,0.3}
\hypersetup{colorlinks,breaklinks,linkcolor=darkblue,urlcolor=darkblue,anchorcolor=darkblue,citecolor=darkblue}

\newcommand{\E}{\mathbb{E}}
\newcommand{\Pp}{\mathbb{P}}

\makeatletter
\def\@setdate{\@date}
\makeatother

\begin{document}
\begin{titlepage}

\title[Misclassification in DID]{Misclassification in Difference-in-Differences Models}

\author{Augustine Denteh\textsuperscript{\dag} \quad  D\'esir\'e K\'edagni\textsuperscript{\ddag}}

\address{\textsuperscript{\dag}Davidson College \quad \textsuperscript{\ddag}UNC-Chapel Hill}

\noindent \date{\scriptsize{First draft on ArXiv:  July 25, 2022. The present version is as of \today. The authors thank Santiago Acerenza, Kyunghoon Ban, Pierre Nguimkeu, Brantly Callaway, Chris Bollinger, and conference participants of the Midwest Econometrics Group, the Southern Economic Association meeting, the Tulane-LSU Applied Microeconomics Conference, and the Africa Meeting of the Econometric Society for helpful discussions and comments. All errors are ours. Augustine Denteh: Davidson College, Department of Economics, 209 Ridge Rd, Davidson, NC 28035, USA. Email address: audenteh@davidson.edu;
D\'esir\'e K\'edagni (Corresponding author): UNC-Chapel Hill, Department of Economics, 204 Gardner Hall, Chapel Hill, NC 27599, USA. Email address: dkedagni@unc.edu}}

\begin{abstract}
This paper studies the identification of the average treatment effect on the treated (ATT) in difference-in-differences (DID) designs when the treatment is misclassified. Misclassification may occur in DID contexts when researchers need to infer treatment from auxiliary data or the timing of a policy intervention is ambiguous. Without imposing any structure on the measurement error, we show that the DID estimand is biased and recovers a non-convex weighted average of ATTs in two subpopulations---the correctly classified and misclassified groups. As such, the DID estimand may yield the wrong sign because the weight for the misclassified group is non-positive. In the case of nondifferential measurement error, we use contextual and data-driven information on the extent of misclassification to bound the ATT, allowing the researcher to assess the sensitivity of their findings. Furthermore, when multiple data sources with varying misclassification rates are available, we construct an identified set for the ATT under nondifferential and arbitrary misclassification, without requiring the researcher to specify misclassification rates. We demonstrate our theoretical results using simulations and two empirical applications to guide researchers in using our proposed methods.
\end{abstract}

\maketitle
\thispagestyle{empty}

\noindent {\footnotesize \textbf{Keywords}: Difference-in-differences, average treatment effect on the treated, misclassification, partial identification, measurement error.

\textbf{JEL subject classification}: C14, C31, C35, C36.}

\end{titlepage}


\section{Introduction}

The difference-in-differences (DID) method is a popular quasi-experimental technique used to identify causal effects of a treatment (e.g., policy intervention) when data is available on the pre- and post-treatment periods. As of 2018, \citet{currie2020technology} reports that 25 percent of National Bureau of Economic Research working papers in applied microeconomics and 15 percent of papers in ``top five'' economics journals mention DID.\footnote{Recently, the DID method has drawn significant attention from methodological researchers working to clarify several identification and estimation issues; see reviews in \citet{deChaisemartin2022} and \citet{roth2022s}.} However, when the treatment variable is observed with errors (in which case we say the treatment is \textit{misclassified}), the DID estimand may not have a clear causal interpretation even when the identifying \textit{parallel trends} (PT) assumption holds.

Misclassification can arise in DID designs from many sources. One common scenario occurs when researchers estimate or infer the treatment variable from auxiliary data, thereby potentially introducing misclassification (\citealp{cortes2013achieving}, \citealp{deChaisemartin2018}, and \citealp{fortson2009hiv}). This scenario frequently occurs when researchers use a proxy variable to classify units into treated and control groups because the treatment variable is missing for either the pre- or post-intervention period. For instance, \citet{groen2008effect} studies the impact of Hurricane Katrina on labor market outcomes of evacuees. The treatment variable in their study is a binary variable for being an evacuee, but the Current Population Survey only collected information on evacuee status after Katrina. As such, the authors defined treated units (evacuees) in the pre-treatment period as those living in Katrina-affected areas, potentially introducing misclassification because not everyone living in those areas evacuated after the storm.

In other cases, researchers use a mismeasured continuous treatment variable to define a binary treatment variable for DID estimation. This is often the case when units are classified as treated when an estimated index or rate exceeds a specified threshold chosen by the researcher (\citealp{Kessler_al2022} and \citealp{Miller2012}).\footnote{Another likely reason for converting continuous treatments to binary variables is the lack of theoretical work on the identification of treatment effects with continuous treatments in DID designs. However, a recent notable exception is \citet{callaway2024difference}.} The resulting binary variable from the mismeasured continuous treatment is necessarily misclassified. Even when the underlying continuous variable is correctly measured, researchers often resort to creating binary variables for DID estimation with the aim of capturing the intensity of exposure to some policy intervention (\citealp{draca2011minimum} and \citealp{galasso2022does}).

In other cases, there is ambiguity regarding the exact timing of a reform's passage or implementation. For instance, when a significant amount of time elapses between a legislation's proposed date and its eventual enactment, the researcher may opt to use the former to define the timing of treatment (\citealp{kresch2020buck}). Researchers might also lack information about the actual implementation of a policy when there is a lag between the passage and its effective implementation or when they use the lag to define the post-intervention period with the aim of allowing sufficient time for the policy's impact to kick in (\citealp{bindler2018punishment} and \citealp{murray2016mice}). Even when researchers know the true treatment date, data unavailability might compel them to define overlapping pre- and post-intervention periods that introduce misclassification (\citealp{buchmueller2011effect}). Most of the above studies admit the misclassification problems confronting them, but there is a conspicuous lack of methodological work addressing it.

In this paper, we study the identification of the average treatment effect on the treated (ATT) in the DID framework when the treatment is subject to misclassification. We characterize the resulting bias and propose a partial identification approach that researchers can use to investigate the sensitivity of their DID estimates. Our framework distinguishes the latent treatment from its observed (misclassified) counterpart and characterizes how the DID estimand aggregates causal effects across correctly classified and misclassified subpopulations. This characterization clarifies when misclassification leads to attenuation, when it can generate sign reversals, and which additional assumptions are needed to recover interpretable causal estimates.

Specifically, we make several contributions to the literature. First, this paper is one of the first to study the identification of causal effects in the DID setting when the treatment is misclassified. In work concurrent to ours, \citet{negi2025difference} apply a one-sided misreporting model (\citealp{nguimkeu2019estimation}) and instrumental variables for both treatment and misreporting to achieve point identification in a parametric DID regression. Their approach is useful when researchers are willing to impose a one-sided misreporting structure and have credible instruments for both treatment and misreporting. In contrast, our contribution targets scenarios where misclassification may be bidirectional, instruments are unavailable, or researchers prefer to avoid strong parametric assumptions. In our framework, we show that under the standard PT assumption, the DID estimand recovers a weighted average of the ATT for the correctly classified and misclassified subpopulations with a non-positive weight for the misclassified units. This finding mirrors the recent ``negative weighting'' criticism of standard two-way fixed effects (TWFE) estimators (\citealp{deChaisemartin2022} and \citealp{goodman2021difference}). However, while the negative weights in those TWFE contexts arise from treatment effect heterogeneity in staggered adoption designs, we demonstrate that misclassification alone can generate non-positive weighting, potentially biasing estimates even in simple canonical DID settings. Furthermore, the weights do not necessarily sum up to one. In general, the direction of the bias is unknown, implying that the DID estimand may induce a sign-reversal phenomenon, where its sign could differ from the true causal effect.

Second, we establish a linear relationship between the ATT and the DID estimand where the coefficients are unidentified. This relationship allows us to discuss conditions under which various types of biases may occur. Using this relationship, we then provide a sufficient condition for the existence of a fixed point where the DID estimand still identifies the ATT even in the presence of misclassification.

Third, we show that under additional assumptions, the DID method identifies the sign of the ATT but remains biased in magnitude. In particular, if the misclassification error is nondifferential and a monotonicity condition holds, then the DID estimand only suffers from attenuation bias, producing smaller estimates in magnitude. When the extent of the misclassification is bounded by a sensitivity parameter, we derive bounds on the ATT under the aforementioned assumptions. The choice of the sensitivity parameter is specific to each application and we suggest that researchers use institutional or contextual knowledge and the structure of their data to inform their choice.

Finally, we develop a partial identification approach to overcome arbitrary misclassification in settings where the researcher has access to multiple data sources drawn from the same underlying population but potentially subject to different rates of misclassification. This scenario arises naturally when, for example, treatment status is recorded in both administrative records and survey data, or when multiple proxy variables are available to classify units into treatment arms. Under the assumption that the data source affects misclassification rates but not potential outcomes or true treatment status, coupled with a relevance condition that misclassification rates vary across sources, we show that the observed outcome distributions within each source decompose into finite mixtures whose components are source-invariant but whose weights vary with the source. We exploit this mixture structure to partially identify stratum-specific ATTs for correctly classified and misclassified subpopulations. We then derive bounds on the unconditional ATT as a convex weighted average of the two stratum-specific ATTs where the weights are partially identified. Unlike the previous results above, these multi-data source bounds are data-driven and do not require the researcher to specify a misclassification rate. We conduct inference using the intersection bounds framework of \citet{CLR2013}.

We illustrate our theoretical results through simulation and empirical exercises. In the simulations, we consider various designs---differential vs nondifferential misclassification, and symmetric vs asymmetric misclassification. The simulation results display sign reversal in the DID estimates when the measurement error is differential regardless of whether the misclassification is symmetric or not. In the case of nondifferential misclassification, we find that the attenuation bias can be substantial depending on the design. For the multi-data source bounds, our simulations show that the method yields informative confidence sets that correctly identify the sign of the ATT.

For applied work, we recommend a simple reporting template that includes the conventional DID estimate under PT and bound sets for the ATT over a transparent range of misclassification rates informed by institutional knowledge. We illustrate these results using two empirical studies. In the main paper, we analyze federalism in the Brazilian water and sanitation sector (\citealp{kresch2020buck}). In Appendix \ref{jury-appendix}, we revisit the impact of abolishing capital punishment in England between 1772 and 1871 (\citealp{bindler2018punishment}). We discuss the possibility that the treatment is misclassified in these studies and illustrate how we can use the data and contextual information to implement our sensitivity bounding analysis with reasonable choices of the sensitivity parameter.

Our work unites the vast literature on difference-in-differences designs and measurement error. Our work is directly related to the longstanding literature on the identification of causal parameters when a binary treatment variable is misclassified. One set of studies in this literature leverages instrumental variables, auxiliary data, or parametric assumptions to achieve point identification (e.g., \citealp{bollinger2017bayesian}; \citealp{nguimkeu2019estimation}). Other studies focus on partial identification and bounding strategies in settings such as heterogeneous instrumental variable models (e.g., \citealp{Acerenza_al2021}; \citealp{Chalak2017}; \citealp{Kreideral2012}; \citealp{Ura2018}; \citealp{possebom2025crime}; \citealp{yanagi2019inference}). Although these studies and the additional papers cited therein cover many quasi-experimental designs, they do not consider misclassification in the DID framework. Our paper also connects with the recent literature exploring a related but different problem of missing data in DID analysis. These studies provide point and partial identification results in the DID framework when the treatment variable is missing for either the pre- or post-treatment period (\citealp{botosaru2018difference} and \citealp{fan2017partial}).

The remainder of the paper is organized as follows. Section \ref{anaF} presents the model, the assumptions, and the main identification results. Section \ref{proposed-bounds} presents our theoretical results for bounding the ATT. Section \ref{MonteCarlo} presents simulation results, and Section~\ref{Empirical} presents an empirical illustration. Section \ref{Conclusion} concludes. Proofs and additional results are in the Appendix.

\section{Analytical Framework}\label{anaF}
Our framework is the canonical DID design comprising two groups and two periods. Consider the following model:\footnote{The specification $D = D^*(1-\varepsilon) + (1-D^*) \varepsilon$ is shown in \citet[Lemma 3]{Acerenza_al2021} to be without loss of generality.}
\begin{eqnarray}\label{seq1}
\left\{ \begin{array}{lcl}
     Y_t&=&Y_t(1)D^*+Y_t(0)(1-D^*),\\ \\
     D &=& D^*(1-\varepsilon) + (1-D^*) \varepsilon
     \end{array} \right.
\end{eqnarray}
where the vector $(Y_0, Y_1,D)$ represents the observed data, while the vector $(Y_t(0), Y_t(1), D^*, \varepsilon)$ is latent. In this model, the variables $Y_t(0)$ and $Y_t(1)$ are the potential outcomes that would have been observed in period $t\in\{0,1\}$ had the treatment been externally set to 0 and 1, respectively.  The variables $Y_0, Y_1 \in \mathcal Y$ are the observed outcomes in the baseline $(t=0)$ and the post-intervention $(t=1)$ periods, respectively. The variable $D^*\in \left\{0,1\right\}$ is the true treatment occurring between periods 0 and 1, while $D$ is a potentially misclassified version of $D^*$. The latent variable $\varepsilon$ is the indicator for misclassification. When  $\varepsilon=0$, there is no misclassification, but the observed treatment is misclassified whenever $\varepsilon=1$.

As is customary in the DID literature, we assume away any anticipatory effects of the treatment, so that $Y_0(1)=Y_0(0)$. We also assume $0< \mathbb P(D^*=1) < 1$, and $0< \mathbb P(D=1) < 1$, implying that a fraction of the population is treated whether or not the observed treatment is misclassified.
In this paper, we are interested in identifying the ATT defined as
\begin{align}\label{trueATT}
	ATT\equiv \mathbb E[Y_1(1)-Y_1(0)\vert D^*=1].
\end{align}

In general, the presence of misclassification complicates the identification of the ATT in model \eqref{seq1}. Such misclassification may induce a violation of the parallel trends assumption in the observed treatment variable $D$, in which case the DID estimand will not have a clear causal interpretation. Even when the parallel trends assumption holds in the misclassified $D$, the causal interpretation of the DID estimand remains unclear. Below, we study the consequences of misclassification on the DID estimand when the researcher assumes parallel trends in $D$. We discuss why the researcher may assume parallel trends in $D$ and provide sufficient conditions for this assumption to hold despite the misclassification. We state the assumptions and present our findings.
\begin{assumption}[Parallel trends with the observed treatment D]\label{PT}
\begin{align*}
\mathbb E\left[ Y_1(0)-Y_0(0) \vert D=1\right] = \mathbb E\left[ Y_1(0)-Y_0(0) \vert D=0\right].
\end{align*}
\end{assumption}
Assumption \ref{PT} is the standard parallel trends assumption commonly used in the literature on difference-in-differences. It states that in the absence of treatment, the control and treatment groups would have followed the same trend on average. It is equivalent to
$$\mathbb E[Y_1(0) \vert D=1]-\mathbb E[Y_1(0) \vert D=0]=\mathbb E[Y_0\vert D=1] - \mathbb E[Y_0 \vert D=0].$$
 In the observed model, the standard DID estimand can be defined as $$\theta_{DID} \equiv \theta^1_{OLS}-\theta^0_{OLS},$$
 where $\theta^0_{OLS}$ and $\theta^1_{OLS}$ are the ordinary least squares (OLS) (or difference-in-means) estimands at periods 0 and 1, respectively given by
 \begin{align*}
 	\theta^1_{OLS} &\equiv \mathbb E[Y_1\vert D=1] - \mathbb E[Y_1 \vert D=0],\\
 	\theta^0_{OLS} &\equiv \mathbb E[Y_0\vert D=1] - \mathbb E[Y_0 \vert D=0].
 \end{align*}

As discussed above, Assumption \ref{PT} is stated in terms of the observed treatment variable (instead of the true, unobserved treatment). This is because when the researcher suspects that the treatment variable is misclassified, they may naturally make Assumption \ref{PT} to proceed with the DID identification strategy for several reasons. They might do so because they want to ignore the problem, (incorrectly) assert that misclassification has minimal consequences or for convenience due to lack of a viable solution (i.e., alternative estimation method). As a result, we study the causal interpretation of the DID estimand under Assumption \ref{PT} for practical considerations. However, before we present the results, we provide sufficient conditions (assumptions) on $(D^*, \varepsilon, Y_1(0), Y_0(0))$ for Assumption \ref{PT} to hold.

\begin{assumption}\label{PTepsDstar} Consider the following pair of parallel trends assumptions with the true treatment $D^*$.
	\begin{enumerate}[(1)]
		\item \label{PTDstar} Parallel trends with the true treatment $D^*$
	\begin{align*}
		\mathbb E\left[ Y_1(0)-Y_0(0) \vert D^*=1\right] = \mathbb E\left[ Y_1(0)-Y_0(0) \vert D^*=0\right].
	\end{align*}

\item \label{PTeps}Parallel trends for correctly classified and misclassified groups within each true treatment arm
	\begin{align*}
	\mathbb E\left[ Y_1(0)-Y_0(0) \vert D^*=d^*, \varepsilon=1\right] = \mathbb E\left[ Y_1(0)-Y_0(0) \vert D^*=d^*, \varepsilon=0\right]\ \text{ for each } d^* \in \{0,1\}.
\end{align*}
\end{enumerate}
\end{assumption}
The first part of Assumption \ref{PTepsDstar} is the standard parallel trends assumption stated in terms of the true treatment variable. The second part states that conditional on the true treatment variable, the average outcomes for the correctly classified and misclassified groups would have followed the same trend. This parallel trends assumption permits different average outcome trends over time across the misclassification groups (defined by $\varepsilon$) in each treatment arm. The following lemma shows that these two assumptions imply Assumption \ref{PT}.
\begin{lemma}\label{lem1}
In model (\ref{seq1}), Assumption \ref{PTepsDstar} implies Assumption \ref{PT}.
\end{lemma}

 \subsection{The DID Estimand under Arbitrary Misclassification}
 This section provides our main results for the consequences of misclassification in the DID framework. We first allow misclassification to be arbitrary (potentially differential) with no structure imposed on it. The following proposition provides an expression of the DID estimand when the misclassified treatment variable is used for identification.
\begin{proposition}\label{prop1}
Suppose that model (\ref{seq1}) along with Assumption \ref{PT} holds. Then, the DID estimand using the misclassified treatment variable can be decomposed as:
\begin{eqnarray}
\theta_{DID} &=& \mathbb E\left[Y_1(1)-Y_1(0) \vert D^*=1, \varepsilon=0\right] \mathbb P(\varepsilon=0\vert D=1) \nonumber\\
&& \qquad - \mathbb E\left[Y_1(1)-Y_1(0) \vert D^*=1, \varepsilon=1\right] \mathbb P(\varepsilon=1\vert D=0). \label{eq:did}
\end{eqnarray}
\end{proposition}
Proposition \ref{prop1} shows that the DID estimand does not recover the true ATT in equation~(\ref{trueATT}), but rather a weighted average of the ATT for two subpopulations---the correctly classified and misclassified treated groups. Importantly, the weights are positive for the correctly observed units and non-positive for the misclassified observations, and do not necessarily sum up to one. It follows that the DID estimand could yield an opposite sign for the treatment effect in some circumstances. Proposition \ref{prop1} shows that the standard DID estimand could be negative even if the true ATT is positive under misclassification and vice versa. Indeed, we note from equation~(\ref{trueATT}) and the law of iterated expectations that
\begin{eqnarray}
ATT &=& \mathbb E\left[Y_1(1)-Y_1(0) \vert D^*=1, \varepsilon=1\right] \mathbb P(\varepsilon=1\vert D^*=1) \nonumber\\
&& \qquad + \mathbb E\left[Y_1(1)-Y_1(0) \vert D^*=1, \varepsilon=0\right] \mathbb P(\varepsilon=0\vert D^*=1). \label{eq:att}
\end{eqnarray}
If the quantities $\mathbb E\left[Y_1(1)-Y_1(0) \vert D^*=1, \varepsilon=0\right]$ and $\mathbb E\left[Y_1(1)-Y_1(0) \vert D^*=1, \varepsilon=1\right]$ are positive, then from equation (\ref{eq:att}), we know that the ATT is positive. It is then straightforward to check that Proposition \ref{prop1} implies the DID estimand could be negative in this scenario. In other words, the DID estimand using a misclassified treatment fails to identify an interesting causal parameter and could potentially yield misleading conclusions. Our pessimistic results suggest that the consequences of misclassification in DID analysis are more severe than previously understood. This finding is contrary to observations in some previous empirical papers where the researchers concerned about possible misclassification suggest that their estimates are potentially only attenuated (\citealp{draca2011minimum}, \citealp{galasso2022does}, and \citealp{Miller2012}).\footnote{For instance, in reference to a binary treatment variable created from a potentially mismeasured continuous variable representing the 2005 county-level uninsurance rate in Massachusetts (Uninsured2005c), \cite{Miller2012} observes that ``One advantage of using a binary indicator, rather than a continuous measure, is that it is not reliant on the assumption of a linear relationship between insurance coverage and emergency room usage and is more robust to measurement error in the variable Uninsured2005c.'' Also, \cite{galasso2022does} remarks that ``Second, because of the threshold approach that we use to define the treatment and control groups, the control subclasses also include implant patents. In principle, this will cause attenuation bias and lead to an underestimation of the impact of the increase in liability.'' }

To elaborate on the nature of the resulting bias due to misclassification, we combine equations (\ref{eq:did}) and (\ref{eq:att}) to obtain the following relationship between the $ATT$ and $\theta_{DID}$.

\begin{proposition}\label{prop2}
Suppose that model (\ref{seq1}) along with Assumption \ref{PT} holds. Then, we have:
\begin{align*}
ATT &= \frac{\mathbb P(D=1)}{\mathbb P(D^*=1)} \theta_{DID} + \frac{\mathbb E[(Y_1(1)-Y_1(0))\varepsilon \vert D^*=1]}{\mathbb P(D=0)}.
\end{align*}
\end{proposition}
Proposition \ref{prop2} shows that the ATT is linearly related to the DID estimand under misclassification, albeit the coefficients are not identified from the observed data. Nonetheless, it is instructive to use this relationship to study the resulting bias.

\begin{figure}
\centering
{\includegraphics[width=0.8\textwidth]{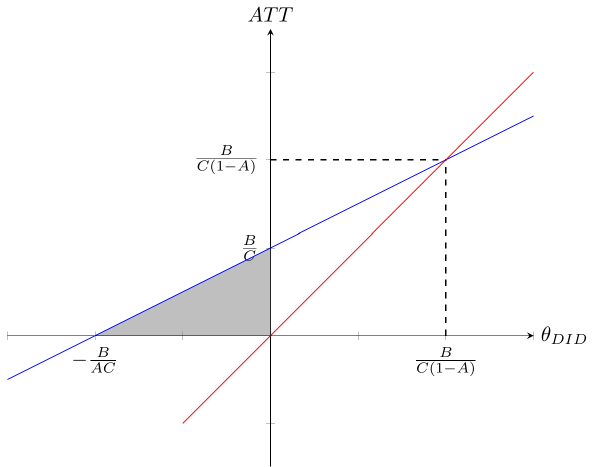} }
\caption{\label{fig1} Graphical illustration of the DID estimand under misclassification.}
\end{figure}

Let the slope coefficient be denoted by $A=\frac{\mathbb P(D=1)}{\mathbb P(D^*=1)}.$ Also, denote the numerator and denominator of the intercept term by $B=\mathbb E[(Y_1(1)-Y_1(0))\varepsilon \vert D^*=1],$ and $C=P(D=0)$, respectively. Furthermore, suppose that the probability of having a false positive is less than that of having a false negative, i.e., $\mathbb P(D^*=0,\varepsilon=1) < \mathbb P(D^*=1,\varepsilon=1)$. This implies $A<1$. Under these assumptions, Figure \ref{fig1} provides a graphical illustration of the bias in the DID estimand under misclassification based on Proposition \ref{prop2}. The blue line is drawn assuming that $B>0$ and the red line represents the 45-degree line.\footnote{For example, $B>0$ could occur when the average treatment effect is positive for the misclassified treated group.} In this case, the sign of the treatment effect is not identified for certain positive values. That is, when the ATT is positive, the DID takes on the wrong (negative) sign whenever $0<ATT<\frac{B}{C}$. The sign-reversal region is indicated by the gray shaded area in Figure \ref{fig1}. Further, when the ATT lies between $\frac{B}{C}$ and $\frac{B}{C(1-A)}$, the DID estimand is biased downwards (attenuation bias). The DID estimand yields an expansion bias whenever $ATT>\frac{B}{C(1-A)}>0$.

To summarize, when $\theta_{DID} < \frac{B}{C(1-A)}$, the DID estimand is biased downwards and it is biased upwards when $\theta_{DID} > \frac{B}{C(1-A)}$. When $\theta_{DID}=\frac{B}{C(1-A)}$, the DID estimand is equal to the ATT, even in the presence of misclassification. This latter scenario corresponds to the fixed point in Figure \ref{fig1}. In Corollary \ref{fixedpoint}, we provide a sufficient condition under which the fixed point result may occur.

We conclude this section with two special cases under arbitrary forms of misclassification. These special cases do not exhaust the possible misclassification mechanisms researchers may be interested in, but they provide results for important practical scenarios. In the first case, we examine one-sided misclassification and assume that only false positives are present. Unidirectional measurement error in a treatment variable has been studied in previous works \citep[e.g.,][]{nguimkeu2019estimation}. An example of one-sided misclassification in DID settings is one discussed in an analysis of the impact of Hurricane Katrina on labor market outcomes of evacuees (\citealp{groen2008effect}). In that study, the treated group was correctly measured in the post-treatment period as those who evacuated due to the storm. However, before the storm, the treated units are subject to misclassification (false positives) because they were determined based on their residence in affected areas. False positives arise because some of those who lived in affected areas before the storm may not have evacuated in the aftermath of the storm, hence not truly treated.

\begin{corollary}\label{nofalseneg}
	Suppose that Assumption \ref{PT} holds, and there are no false negatives, i.e., $\mathbb P(D^*=1, \varepsilon=1)=0$. Then,
	\begin{align*}
		\theta_{DID} &= \mathbb P(\varepsilon=0 \vert D=1) ATT\ \ \text{  (attenuation bias)}
	\end{align*}
\end{corollary}

Corollary \ref{nofalseneg} shows that even when misclassification is arbitrary, its consequence for DID estimation is less severe when  the misclassification is one-sided. In the next section, we provide alternative conditions under which attenuation bias can occur.

In the second special case, we consider what happens when misclassification arises following a Roy selection mechanism such that units are misclassified when their treatment effect exceeds some threshold.

\begin{assumption}[Roy selection misclassification]\label{Roymis}
	$\varepsilon=\mathbbm{1}\{Y_1(1)-Y_1(0) >q\}$
\end{assumption}
\begin{corollary}\label{CorRoymis}
	Suppose that Assumptions \ref{PT} and \ref{Roymis} hold. Then, the following inequality holds:
	\begin{align*}
		ATT \geq \frac{\mathbb P(D=1)}{\mathbb P(D^*=1)} \theta_{DID}+q \frac{\mathbb P(\varepsilon=1 \vert D^*=1)}{\mathbb P(D=0)}.
	\end{align*}
\end{corollary}

Corollary \ref{CorRoymis} implies that if $q$ is positive and the DID estimand $\theta_{DID}$ is also positive, the sign of the ATT is identified as positive. But, it is also possible that $\theta_{DID}$ be negative while the $ATT$ is positive. When the threshold $q$ is ``sufficiently'' large, we can ensure that the $ATT$ is positive regardless of the value of $\theta_{DID}$. The proof of this corollary follows from Proposition \ref{prop2} and Assumption \ref{Roymis}.

\subsection{The DID Estimand under Nondifferential Misclassification}
We now show how the DID estimand fares under additional assumptions on the nature of misclassification. The literature on measurement error often invokes additional assumptions when the researcher believes that the nature of misclassification is uncorrelated with potential outcomes.\footnote{See \cite{bound2001measurement} for additional discussions on various types of measurement errors.} We show that if the measurement error is nondifferential and a monotonicity condition holds, then the DID estimand yields an attenuation bias.

\begin{assumption}[Nondifferential misclassification]\label{Nondiff}
$\varepsilon\ \indep\ \left(Y_1(1),Y_1(0)\right) \vert D^*$
\end{assumption}
Assumption \ref{Nondiff} states that conditional on the true (unobserved) treatment, misclassification is independent of the potential outcomes. This type of measurement error is likely in some empirical contexts, especially when the misclassification stems from uncertainty regarding the timing of treatment (e.g., when the researcher uses a reform's proposed date instead of its date of passage).

\begin{assumption}[Monotonicity condition]\label{Mon}
$
\mathbb P(\varepsilon=1\vert D=1)+\mathbb P(\varepsilon=1\vert D=0) < 1.
$
\end{assumption}
Assumption \ref{Mon} is equivalent to the well-known monotonicity condition in \cite{Hausman_al1998}, which states that the sum of false positive and false negative rates of misclassification may not exceed one, i.e., $\mathbb P(\varepsilon=1\vert D^*=1)+\mathbb P(\varepsilon=1\vert D^*=0) < 1$. In Appendix~\ref{apx:mon_equivalence}, we prove the equivalence between Assumption \ref{Mon} and the monotonicity condition in  \cite{Hausman_al1998}. Assumption \ref{Mon} is a minimal requirement to ensure that the misclassification problem is not too severe to render the research project infeasible.

Under Assumption \ref{Nondiff}, we have $\mathbb E\left[Y_1(1)-Y_1(0) \vert D^*, \varepsilon \right]=\mathbb E\left[Y_1(1)-Y_1(0) \vert D^*\right]$. Hence, the following corollary holds.
\begin{corollary}\label{attenuate}
Under Assumptions \ref{PT}, \ref{Nondiff} and \ref{Mon}, we have:
\begin{eqnarray}
\theta_{DID} &=& ATT \left(1-\mathbb P(\varepsilon=1\vert D=1)-\mathbb P(\varepsilon=1\vert D=0)\right),
\end{eqnarray}
where $1-\mathbb P(\varepsilon=1\vert D=1)-\mathbb P(\varepsilon=1\vert D=0) \in (0,1].$
\end{corollary}
Corollary \ref{attenuate} shows that the standard DID method under Assumptions \ref{PT}, \ref{Nondiff} and \ref{Mon} produces smaller effects in magnitude when the treatment is misclassified (attenuation bias).

As mentioned earlier, we now provide a sufficient condition that yields a fixed point result under nondifferential misclassification. Here, the DID method is theoretically unaffected by the presence of misclassification.
\begin{corollary}\label{fixedpoint}
	Suppose that Assumptions \ref{PT} and \ref{Nondiff} hold. Suppose also that $\mathbb P(D^*=1)\neq \mathbb P(D=1)$. Whenever $1-\mathbb P(\varepsilon=1\vert D=1)-\mathbb P(\varepsilon=1\vert D=0)=\frac{\mathbb P(\varepsilon=1, D^*=1)}{\mathbb P(D=0) \left(\mathbb P(D^*=1, \varepsilon=1)-\mathbb P(D^*=0, \varepsilon=1)\right)}$, then $ATT=\theta_{DID}$, and the presence of misclassification in the treatment variable will not induce bias in the DID estimand.
\end{corollary}

Corollary \ref{fixedpoint} provides a sufficient condition under which there exists a fixed point in the relationship between the $ATT$ and the DID estimand. At the fixed point, the DID estimand is robust to any misclassification in the treatment variable.

\section{Bounding the treatment effect using additional information}\label{proposed-bounds}
Given the above results, how can researchers overcome the bias from misclassification? In the measurement error literature, there are various approaches to addressing misclassification, including partial and point identification methods. We believe that the appropriate approach for overcoming measurement error in the DID framework is context-dependent, especially given the various ways misclassification may arise. If the researcher is willing to impose a one-sided measurement error structure and maintain parametric assumptions about the error structure, \cite{negi2025difference} propose a two-step estimator. In the next two subsections, we propose complementary partial identification methods for addressing misclassification under nondifferential and arbitrary misclassification.

\subsection{Nondifferential Misclassification}
If the researcher is willing to invoke the assumption of nondifferential misclassification (Assumption \ref{Nondiff}), we develop bounds on the ATT that allow them to investigate the sensitivity of their results to misclassification using contextual knowledge to infer the extent of misclassification. Specifically, we use the extent of misclassification to bound the ATT by scaling the DID estimate accordingly. We, therefore, introduce the following assumption and derive the bounds in the subsequent corollary.

\begin{assumption}[Known upper bound on the extent of misclassification]\label{Upperbound}
$$\mathbb P(\varepsilon=1\vert D=1)+\mathbb P(\varepsilon=1\vert D=0) \leq \lambda < 1.$$
\end{assumption}
\begin{corollary}\label{bounds}
Under Assumptions \ref{PT}, \ref{Nondiff}, \ref{Mon}, and \ref{Upperbound}, we have:
\begin{eqnarray}
\min\left\{\frac{\theta_{DID}}{1-\lambda}, \theta_{DID}\right\} \leq ATT \leq  \max\left\{\frac{\theta_{DID}}{1-\lambda}, \theta_{DID}\right\}.
\end{eqnarray}
\end{corollary}
Corollary \ref{bounds} provides bounds on the ATT based on the DID estimand and the upper bound on the extent of misclassification, $\lambda$, which plays the role of a sensitivity parameter. Larger values of $\lambda$ imply wider bounds for the ATT, while smaller values of $\lambda$ yield tighter bounds. In the special case where $\lambda=0$ (no misclassification), the bounds collapse to a point, and the ATT is point-identified as the standard DID estimand. In the next subsection, we develop an alternative partial identification approach to bounding the ATT if the researcher departs from Assumption \ref{Nondiff} and has access to multiple data sources drawn from the same population.

\subsection{Arbitrary Misclassification}

Under arbitrary forms of misclassification, we show how to combine multiple exogenous data sources to partially identify the treatment effect. We introduce the following identifying assumptions.

\begin{assumption}[Data source exogeneity]\label{ass:datasourceexogeneity}
(i) For each $d^*\in \{0,1\},$ $Y_t(d^*,s)=Y_t(d^*,s')$, $D^*(s)=D^*(s'),$ for all $s,s' \in \mathcal S,$ and (ii) $S \indep (Y_t(0,s), Y_t(1,s), D^*(s)) \vert \varepsilon,$ where $Y_t(d^*,s)$ is the potential outcome in period $t$ when the true treatment and the data source are externally set to $d^*$ and $s$, respectively, and $D^*(s)$ is the potential true treatment when the data source is externally set to $s$.
\end{assumption}
Assumption \ref{ass:datasourceexogeneity} states that $(i)$ the data source is excluded from the outcome and the actual treatment status, and $(ii)$ the potential outcomes and the actual treatment status are independent of the data source conditional on misclassification. However, note that data source $S$ is allowed to influence misclassification $\varepsilon$. This assumption is intuitive and is likely to hold in many settings because survey administrators who collect the data and policymakers are typically separate.

\begin{assumption}[Relevance]\label{ass:relevance}
For each observed treatment arm $d$, there exist $s_0^d, s_1^d \in \mathcal S$ such that $ \mathbb P(\varepsilon=1 \vert D=d, S=s_0^d) \neq \mathbb P(\varepsilon=1 \vert D=d, S=s_1^d).$
\end{assumption}
Assumption \ref{ass:relevance} states that the misclassification probability for each observed treatment status is changing with the data source. This assumption combined with Assumption \ref{ass:datasourceexogeneity} imply that the data source induces some variations in the observed outcome and treatment distributions without changing the potential outcome and potential treatment distributions.

Under Assumption \ref{ass:datasourceexogeneity}, we show in Appendix \ref{combinedata} that the following equalities hold.
\begin{eqnarray*}
\mathbb P(Y_t \leq y \vert D=1, S=s)
&=&\mathbb P(Y_t(0) \leq y\vert  \varepsilon=1, D^*=0)  \mathbb P(\varepsilon=1\vert D=1, S=s) \\
&& \qquad \qquad +\ \mathbb P(Y_t(1) \leq y\vert  \varepsilon=0, D^*=1)  \mathbb P(\varepsilon=0\vert D=1, S=s),\\
\mathbb P(Y_t \leq y \vert D=0, S=s)&=&\mathbb P(Y_t(1) \leq y\vert  \varepsilon=1, D^*=1)  \mathbb P(\varepsilon=1\vert D=0, S=s) \\
&& \qquad \qquad +\ \mathbb P(Y_t(0) \leq y\vert  \varepsilon=0, D^*=0)  \mathbb P(\varepsilon=0\vert D=0, S=s).
\end{eqnarray*}
We have a finite mixture model where mixture weights vary across data sources while mixture distributions do not. We follow \cite{HKS2014} and \cite{kedagni2023identifying} to partially identify the mixture components. See details in Appendix \ref{combinedata}.

To partially identify the ATT, we rely on the following standard parallel trends assumption (stated in terms of the true treatment) for the correctly classified and misclassified groups.
\begin{assumption}\label{PTdstarallgroups}
$\mathbb E[Y_1(0)-Y_0(0) \vert \varepsilon=\ell,D^*=1]=\mathbb E[Y_1(0)-Y_0(0) \vert \varepsilon=\ell,D^*=0]$ for all $\ell \in \{0,1\}.$
\end{assumption}
In the following proposition, we use the notation $\alpha^d(s)\equiv \mathbb P(\varepsilon=1\vert D=d, S=s)$.
\begin{proposition}\label{prop:ATT_uncond}
		Suppose that model \eqref{seq1} and Assumptions \ref{ass:datasourceexogeneity}-\ref{PTdstarallgroups} hold. Let $\mathcal A_0$ and $\mathcal A_1$ denote the identified
		sets for $ATT_{\varepsilon=0}$ and $ATT_{\varepsilon=1}$, respectively, implied by Proposition~\ref{prop:idATTepsilons} (Appendix). Let $p \equiv \Pp(\varepsilon=1\mid D^*=1).$
		Then the unconditional ATT satisfies
		\begin{align}\label{eq:ATT_uncond_decomp}
			ATT = (1-p)\,ATT_{\varepsilon=0} + p\,ATT_{\varepsilon=1},
		\end{align}
		and the identified set for $ATT$ is
		\[
		\mathcal A \equiv \Big\{(1-p)a_0 + p a_1:\ a_0\in\mathcal A_0,\ a_1\in\mathcal A_1,\ p\in\mathcal P\Big\},
		\]
		where $\mathcal P$ is the identified set for $p$ implied by the (partially identified) weight functions $\alpha^1(\cdot)$ and $\alpha^0(\cdot)$
		and the observed distribution of $(D,S)$.

		In particular, if $S$ is observed in a pooled dataset and $\Pp(S=s)$ and $\pi(s)\equiv \Pp(D=1\mid S=s)$ are known, then
		\begin{align}\label{eq:p_formula}
			p
			=
			\frac{\sum_{s\in\mathcal S}\alpha^0(s)\big(1-\pi(s)\big)\Pp(S=s)}
			{\sum_{s\in\mathcal S}\left[(1-\alpha^1(s))\pi(s)+\alpha^0(s)\big(1-\pi(s)\big)\right]\Pp(S=s)},
		\end{align}
		so $\mathcal P$ is obtained by allowing $(\alpha^1(\cdot),\alpha^0(\cdot))$ to range over values consistent with
		Theorem \ref{thm:id1} (Appendix) and then mapping them through equation \eqref{eq:p_formula}.
\end{proposition}

\begin{proposition}\label{prop:ATT_uncond_prime}
Suppose that model \eqref{seq1}, Assumptions \ref{Nondiff} and \ref{ass:datasourceexogeneity}-\ref{PTdstarallgroups} hold. Let $\mathcal A_0$ and $\mathcal A_1$
denote the identified sets for $ATT_{\varepsilon=0}$ and $ATT_{\varepsilon=1}$, respectively, implied by Proposition \ref{prop:idATTepsilons} (Appendix).
Then
\[
ATT_{\varepsilon=0}=ATT_{\varepsilon=1}=ATT,
\]
and the identified set for the unconditional $ATT$ tightens to
\[
\mathcal A^{ND}\equiv \mathcal A_0\cap \mathcal A_1.
\]
In particular, $\mathcal A^{ND}\subseteq \mathcal A$ (where $\mathcal A$ is defined in Proposition \ref{prop:ATT_uncond}), and the mixing weight $p\equiv \Pp(\varepsilon=1\mid D^*=1)$ is not required to bound $ATT$ under Assumption \ref{Nondiff}.
\end{proposition}

\section{Monte Carlo study}\label{MonteCarlo}
In this section, we present Monte Carlo simulations to illustrate the consequences of misclassification in DID designs and investigate the usefulness of our proposed bounds. Our simulations cover several data generation processes (DGPs) corresponding to various types (differential and nondifferential) and degrees of misclassification.

\subsection{Simulation design.} All our simulation designs have the same basic structure but differ in terms of the nature of misclassification. We draw potential outcomes based on the two-period model in (\ref{seq1}) as follows:
	\begin{align*}
		Y_{0}(0) &\sim D^* \mathcal U_{[-6,0]} + (1-D^*) \mathcal U_{[0,2]} +     u_0, \\
		Y_{1}(1)  &\sim D^* \mathcal U_{[-3,3]} + (1-D^*) \mathcal U_{[3,5]} + u_1, \\
		Y_1(0)&= Y_0(0),
	\end{align*}
where $Y_{0}(0)$ and $Y_{1}(d), d\in \{0,1\},$ are the pre-treatment and post-intervention period potential outcomes, respectively; $\mathcal U_{[a,b]}$ are continuous uniform random variables on the interval $[a,b]$; $u_j, j\in \{0,1\},$ are standard normal random variables; and $D^*$ is the true, unobserved treatment. To focus on the misclassification problem, all our designs maintain the parallel trends assumption by setting $Y_1(0)=Y_0(0)$.

We generate the true treatment indicators as Bernoulli random variables with a success probability of 0.5. The researcher observes a possibly misclassified treatment variable depending on the misclassification dummy, $\varepsilon$. When misclassification is differential, we generate the misclassification dummy as $\varepsilon = \mathbbm{1}\{Y_1(1)-Y_1(0) >q\}$, where $q$ denotes an appropriately chosen quantile of the distribution of $Y_1(1)-Y_1(0)$ to obtain the desired level of misclassification. For nondifferential misclassification, $ \varepsilon = D^* \mathcal Bernoulli(p)+ (1-D^*)(1-\mathcal Bernoulli(p)),$ where the success probability, $p,$ determines the degree of misclassification.

In both cases (differential and nondifferential), we consider various types of misclassification. We examine the scenario where the misclassification is symmetric or asymmetric across the true treatment states. We refer to misclassification as symmetric when we have equal rates of false negatives and false positives; the errors are asymmetric when those error rates are different. We also study the special case where the misclassification is one-sided, with only false positives or false negatives being present.

For the multi-data source simulations, the nondifferential case uses the same Bernoulli mechanism described above, while the differential case uses an alternative DGP in which the misclassification probability depends on how extreme the individual treatment effect (ITE) gains are. Specifically, let $ITE = Y_{1}(1) - Y_{1}(0)$ denote the individual treatment effect, and define the standardized score $Z = |ITE - ATE|/\sigma_{ITE}$, where $ATE=\mathbb E[ITE]$ and $\sigma_{ITE}$ are the cross-sectional expectation and standard deviation of $ITE$. Individuals with the highest scores are deterministically misclassified: the misclassification indicator is $\varepsilon = \mathbbm{1}\{Z \geq z^*\}$, where $z^*$ is the threshold that yields the desired misclassification rate. We consider the case where each data source $s$ has a different misclassification rate (i.e., 5\% vs. 10\%).

The observed data is given by $\{Y_{0}, Y_{1}, D \}$, where the outcomes (pre- and post-treatment) and treatment variable are respectively governed by the following observation mechanism. For outcomes,  $Y_0 = Y_{0}(0)$ and $Y_1=Y_1(1)D^*+Y_1(0)(1-D^*)$, and for treatment,  $D = D^*(1-\varepsilon) + (1-D^*) \varepsilon$.

\subsection{Simulation results.}
This section presents our Monte Carlo study results based on samples of size 10,000, which are aggregated across 10,000 iterations. The true treatment effect (ATT) equals~3 in all the experiments. Overall, the simulation results show that DID estimator is severely biased downwards and sometimes yields a sign opposite of the true treatment effect. This general finding aligns with our main theoretical result in Proposition \ref{prop1}. Even when the parallel trends assumption holds, we cannot generally sign the ensuing bias in DID estimation when misclassification is allowed to be differential.\footnote{Appendix Table \ref{tab2} presents simulation results for nondifferential misclassification. Those results show that the DID estimates exhibit an attenuation bias as shown in Corollary \ref{attenuate} for all levels of misclassification. In addition, the proposed bounds in Corollary \ref{bounds} are reported in Column 5 of Appendix Table \ref{tab2}.}

\begin{table}[htp!]
	\centering
	\caption{\textbf{Simulation Results for Differential Misclassification of Treatment}}
	\label{tab1}
	\begin{threeparttable}
\begin{tabular}{@{}ccccc@{}}
\toprule
\multirow[c]{2}{*}{\textbf{Overall error rate}} & \multirow[c]{2}{*}{\parbox{2cm}{\centering \textbf{False negative}}} & \multirow[c]{2}{*}{\parbox{2cm}{\centering \textbf{False positive}}} & \multicolumn{2}{c}{\textbf{DID estimates}} \\
\cmidrule{4-5}      &       &       & \textbf{True treatment} & \textbf{Observed treatment} \\
\addlinespace
& (1) &(2) & (3) &  (4) \\
\midrule
\multicolumn{3}{l}{\underline{Panel A: Symmetric errors}} &       &                \\
5\%   & 0.050 & 0.050 & 3.000 & 2.134 \\
10\%   & 0.100 & 0.100 & 3.000 & 1.419 \\
20\%   & 0.200 & 0.200 & 3.000 & 0.209 \\
30\%   & 0.300 & 0.300 & 3.000 & \textcolor{black}{-0.789} \\
40\%   & 0.400 & 0.400 & 3.000 & \textcolor{black}{-1.617} \\
50\%   & 0.500 & 0.500 & 3.000 & \textcolor{black}{-2.290} \\
 \hline 
\addlinespace
\multicolumn{3}{l}{\underline{Panel B: Asymmetric errors}} &       &                \\
5\%   & 0.083 & 0.011 & 3.000 & 1.807 \\
10\%   & 0.155 & 0.024 & 3.000 & 0.937 \\
20\%   & 0.273 & 0.059 & 3.000 & \textcolor{black}{-0.384} \\
30\%   & 0.365 & 0.115 & 3.000 & \textcolor{black}{-1.403} \\
40\%   & 0.439 & 0.222 & 3.000 & \textcolor{black}{-2.242} \\
50\%   & 0.486 & 0.396 & 3.000 & \textcolor{black}{-2.688} \\
 \hline 
\addlinespace
\multicolumn{3}{l}{\underline{Panel C: False negatives only}} &       &                \\
5\%   & 0.084 & 0.000 & 3.000 & 1.829 \\
10\%   & 0.179 & 0.000 & 2.999 & 0.698 \\
20\%   & 0.281 & 0.000 & 3.000 & \textcolor{black}{-0.443} \\
30\%   & 0.393 & 0.000 & 3.000 & \textcolor{black}{-1.846} \\
40\%   & 0.456 & 0.000 & 3.000 & \textcolor{black}{-3.030} \\
50\%   & 0.495 & 0.000 & 3.000 & \textcolor{black}{-5.023} \\
 \hline 
\addlinespace
\multicolumn{3}{l}{\underline{Panel D: False positives only}} &       &                \\
5\%   & 0.000 & 0.075 & 3.001 & 2.775 \\
10\%   & 0.000 & 0.199 & 3.000 & 2.404 \\
20\%   & 0.000 & 0.304 & 3.001 & 2.090 \\
30\%   & 0.000 & 0.386 & 3.000 & 1.843 \\
40\%   & 0.000 & 0.448 & 2.999 & 1.656 \\
50\%   & 0.000 & 0.498 & 3.000 & 1.506 \\
\bottomrule 
\end{tabular}%

		\begin{tablenotes}[flushleft]
			\footnotesize
            \setlength{\baselineskip}{11pt}
			\item \textbf{Notes:} This table presents simulation results for the case of differential misclassification of treatment within the differences-in-differences framework. Columns 1 and 2 report false negative and false positive rates conditional on true treatment status. The DID estimates using the true (unobserved) and misclassified (observed) treatment variables are reported in Columns 3 and 4, respectively. The panels correspond to various types of misclassification discussed in the Monte Carlo design setup section.
		\end{tablenotes}
	\end{threeparttable}
\end{table}

\begin{table}[htp!]
\centering
\caption{\textbf{Bounds on the ATT Using Multiple Data Sources}}
\label{tab:multisource}
\begin{threeparttable}
\begin{tabular}{lcc}
\toprule
 & Proposition 3.1 & Proposition 3.2 \\
 & (Arbitrary misclassification) & (Nondifferential) \\
\midrule
\multicolumn{3}{l}{\textit{Panel A: Nondifferential misclassification}} \\[3pt]
95\% CI & [$0.864$,\ $4.195$] & [$0.864$,\ $4.193$] \\[6pt]
\midrule
\multicolumn{3}{l}{\textit{Panel B: Differential misclassification}} \\[3pt]
95\% CI & [$0.905$,\ $4.280$] & [$0.911$,\ $4.273$] \\[6pt]
\bottomrule
\end{tabular}
\begin{tablenotes}[flushleft]
\footnotesize
\setlength{\baselineskip}{11pt}
\item \textbf{Notes:} This table presents 95\% confidence intervals for bounds on the ATT using two data sources with different misclassification rates (5\% vs.\ 10\%). Panel~A uses a nondifferential misclassification DGP and Panel~B uses a differential misclassification DGP and the estimates are averaged across 100 iterations. For Panel A, the average standard DID for the 5\% and 10\% data sources are 2.703 (0.107) and 2.411 (0.111), respectively. For Panel B, the average standard DID for the 5\% and 10\% data sources are 2.731 (0.110) and 2.459 (0.115), respectively. 
\end{tablenotes}
\end{threeparttable}
\end{table}

For differential misclassification (Table \ref{tab1}), the DID estimates have the wrong (negative) signs in all instances (Panels A through C) except for the case where only false positives are present. We find that the sign switching of the DID estimate occurs with at least a 20\% unconditional misclassification rate in Panels B and C but with at least 30\% in Panel A. All else equal, one likely reason why the sign reversal occurs at higher levels of misclassification in Panel A is that the errors are symmetric and the biases (from false negatives and false positives) might cancel out more evenly, leading to a less severe overall bias. The sign-reversal result we obtain in the DID context when the parallel trends assumption holds bears resemblance to the consequences of measurement error in linear treatment effect models under strict exogeneity. For instance, \citet{nguimkeu2019estimation} shows that, even when treatment is exogenous, the sign of the OLS estimator is not generally identified with endogenous (differential) misreporting.

In the case where the misclassification is solely false positives (Panel D), we find that the DID estimates are attenuated but take on the correct (positive) sign of the true ATT. This finding is consistent with Proposition \ref{prop1} by observing that the second term on the right-hand side of equation \eqref{eq:did} is zero when there are no false negatives. The simulation results illustrate the severe consequences of misclassification, but do not exhaust the set of possible outcomes thereof. As shown in Proposition \ref{prop2} and the discussion immediately following it, the resulting bias in the DID estimator can take any form---attenuation bias, expansion bias, or sign reversal---depending on various parameters in specific contexts and data generating processes. In summary, the simulations illustrate our theoretical results showing that the consequences of misclassification in DID designs can be severe, sometimes going beyond an attenuation bias to producing the incorrect sign of the treatment effect.

Table~\ref{tab:multisource} presents simulation-based confidence intervals for the ATT bounds derived from combining two data sources with different misclassification rates. For both the nondifferential case (Panel~A) and the differential case (Panel~B), the 95\% confidence intervals based on Propositions~\ref{prop:ATT_uncond} and~\ref{prop:ATT_uncond_prime} correctly identify the true sign of the ATT, demonstrating that exploiting variation in misclassification rates across data sources yields informative bounds. The multi-data source theoretical results and simulations suggest this partial identification approach may be a viable solution to measurement error if the researcher can access multiple data sources applicable to their research question.

\section{Empirical Illustration}\label{Empirical}

In this section, we illustrate our theoretical results using an empirical application highlighting how misclassification may arise in DID designs. Our objective is to demonstrate the usefulness of our proposed methods and provide guidance on how researchers may use them. The application considers the effect of a 2007 legal reform in Brazil that legislated municipal governments (as opposed to state governments) as the ultimate authority to provide public services in the water and sanitation sector \citep{kresch2020buck}. A second empirical illustration revisiting how punishment severity affects jury decision-making \citep{bindler2018punishment} is provided in Appendix~\ref{jury-appendix}.

\subsection{Federalism in the Brazilian water and sanitation sector}
In January 2007, the Brazilian Congress passed a law (National Water Law 11.447) that gave municipalities the ultimate authority to provide water and sanitation services, eliminating the risk of state governments overtaking municipal-run companies. \citet{kresch2020buck} uses a DID design comparing self-run municipalities (treatment) to state-run municipalities (control) and finds that this reform led to significant increases in total system investment, access, and decreases in child mortality.\footnote{As in the original study, we maintain that the parallel trends assumption plausibly holds given that the decision to be a self-run or state-run municipality was made in the 1970s, with no companies switching their status during the study period.}

Crucially, \citet{kresch2020buck} used the date of the proposed legislation (2005) rather than its passage date (2007) to define the post-intervention period, arguing that the bill's likely passage may have shifted investment decisions earlier. We view this choice as introducing potential misclassification in the DID framework.\footnote{One alternative interpretation is that using the proposed date identifies \textit{anticipatory effects} of the reform rather than introducing misclassification. However, we think this alternate view moves the goalpost and changes the interpretation of the policy-relevant parameter identified by the DID estimand.} Unlike most empirical settings, \citet{kresch2020buck} can directly compare estimates using both dates. Our proposed bounds provide a complementary sensitivity tool for researchers who lack this advantage.

Table \ref{tab3} presents our findings on the effect of the legal reform on investment decisions. We replicate the original analysis without covariates.\footnote{Our results replicate the estimates in Table 3 of the original study, albeit without covariates. Some of the covariates used in the original study include municipality characteristics such as population size, municipality finance measures, and temperature and rainfall variables.} Columns 1 and 2 present the DID estimates using 2005 and 2007 to mark the post-intervention periods, respectively. Both sets of DID estimates show that the legal reform eliminating take-over risk led to a statistically significant increase in overall investments and all types of investments except for government grants and investment in water. While the results across the two legislation dates are qualitatively the same, the estimates using the earlier proposed date are slightly smaller.\footnote{However, we cannot reject the null hypothesis that both sets of estimates are equal at the 5\% level of significance.}

\begin{table}[htp!]
	\centering
	\caption{\textbf{The Impact of Water and Sanitation Legal Reform on Investment}}
	\label{tab3}
	\begin{threeparttable}
		\begin{tabular}{@{}lccc@{}}
\toprule
\textbf{Dependent variables} & \parbox{2.5cm}{\centering \textbf{Proposed date (2005)}} & \parbox{2.2cm}{\centering \textbf{Passage date (2007)}} & \parbox{2cm}{\centering \textbf{ATT bounds ($\lambda=0.29$)}} \\
\addlinespace
& (1) &(2) & (3)  \\
\midrule
\emph{Panel A: Overall investment}&  & &  \\
\addlinespace
Total Investment & 3,226.26$^{**}$  & 3,287.98$^{*}$  & (3,226.26 , 4,516.77) \\
& (1,423.34) & (1,695.24) &  \\
\addlinespace
\emph{Panel B: Sources of investments}&  & &  \\
\addlinespace
Self-financing & 1,835.43$^{***}$  & 1,930.41$^{***}$  & (1,835.43 , 2,569.60) \\
& (481.52) & (520.57) &  \\
\addlinespace
Loans and Debt & 2,121.43$^{**}$  & 2,273.94$^{**}$  & (2,121.43 , 2,970.00) \\
& (897.80) & (1,003.09) &  \\
\addlinespace
Government grants & 36.17  & 40.20  & (36.17 , 50.64) \\
& (223.99) & (296.62) &  \\
\addlinespace
\emph{Panel C: Destination of investments}&  & &  \\
\addlinespace
Investment in Water & 869.50  & 787.78  & (869.50 , 1,217.29) \\
& (536.52) & (689.35) &  \\
\addlinespace
Investment in Sewer & 1,894.92$^{**}$  & 2,025.37$^{*}$  & (1,894.92 , 2,652.89) \\
& (948.32) & (1,178.14) &  \\
\addlinespace
Other Investments & 450.52$^{***}$  & 501.72$^{***}$  & (450.52 , 630.72) \\
& (146.63) & (175.80) &  \\
\addlinespace
\hline
Observations &       14,460 &       14,460 &  \\

\bottomrule
\end{tabular}%

		\begin{tablenotes}[flushleft]
			\footnotesize
            \setlength{\baselineskip}{11pt}
			\item \textbf{Notes:} This table presents difference-in-differences estimates of the impact of the 2007 legal reform in the Brazilian water and sanitation sector on investment decisions based on the analysis sample in \cite{kresch2020buck}. The sample comprises yearly municipality-level data on the total system investments (grouped according to the origin of funds) from 2001 to 2012. All analyses are conducted without covariates. Investment amounts are measured in thousand reals. The DID estimates in column (1) use the legislation's proposed date as in Kresch's original analysis (2005) to demarcate the pre- and post-periods while those in column (2) are based on the legislation's passage date (2007). The bounds on the ATT in Column (3) are point-estimate bounds based on Corollary \ref{bounds} accounting for the misclassification resulting from the choice of the post-treatment period.
		\end{tablenotes}
	\end{threeparttable}
\end{table}

Column 3 presents bounds on the ATT based on Corollary \ref{bounds} under nondifferential misclassification. We estimate the sensitivity parameter as $\lambda=0.29$ (i.e., $2/7$), reflecting the two-year discrepancy between the proposed and passage dates relative to the seven-year post-intervention period.\footnote{Alternatively, one can report alternative bounds using various rates of misclassification based on reasonable beliefs regarding the extent of misclassification.} The upper limits of the bounds are roughly 40\% higher than the DID estimates (based on the proposal date) across most investment types, and in all but one instance the bounds include the DID estimate using the passage date. This finding is reassuring: our proposed bounds provide an informative and complementary way to assess the sensitivity of DID estimates to misclassification when the researcher cannot directly verify robustness to alternative treatment definitions.

\section{Conclusion}\label{Conclusion}

The difference-in-differences design remains one of the most widely used quasi-experimental designs in economics and related disciplines. Under a parallel trends assumption, this method identifies the average treated effect on the treated. Part of the DID's appeal is its simplicity and ease of use. The canonical DID design provides a nonparametric estimate of the ATT by comparing the differences in outcomes across two groups and periods. The DID method continues to attract a great deal of attention from methodological researchers along several dimensions. One aspect of the DID design that has received little to no attention is the misclassification of treatment status. When empirical researchers encounter misclassification in their DID analysis, they often assert that its consequences are likely minimal, resulting in an attenuation bias.

This paper investigates the identification of the DID estimand under arbitrary misclassification. We show that the consequences of misclassification are more severe than previously articulated in the literature, with the sign of the DID estimand possibly being different from the true treatment effect. In particular, the DID estimand using a possibly misclassified treatment variable identifies a weighted average of the ATT for the correctly classified and misclassified subgroups, with the weights being negative for the latter.

Under nondifferential misclassification, we find that the DID estimand is attenuated but recovers the correct sign. We propose bounds on the ATT under reasonable assumptions on the extent of misclassification. Researchers may use these bounds for sensitivity analysis when they suspect misclassification of the treatment variable in their DID framework.
We also propose a partial identification approach under arbitrary misclassification when the researcher has access to multiple data sources drawn from the same underlying population but subject to different misclassification rates.

We provide simulation evidence to demonstrate our theoretical results and conclude with empirical applications that provide guidance on how to estimate the sensitivity parameter used for our bounding exercise. This paper contributes to the literature by providing new insights on the implications of measurement error in DID designs. Future work can consider extensions of our work to staggered DID designs and consider the inclusion of covariates.

\clearpage
\bibliographystyle{chicago}
\bibliography{references}


\appendix
\renewcommand\thetable{\Alph{section}.\arabic{table}}
\renewcommand\thefigure{\Alph{section}.\arabic{figure}}

\clearpage
\section{Proofs of the main results}\label{proofs-appendix}

\subsection{Proof of Lemma \ref{lem1}}
We have
\begin{eqnarray*}
	\mathbb E[Y_t(0)\vert D=1]-\mathbb E[Y_t(0) \vert D=0] &= \mathbb E[Y_t(0)\vert D=1, D^*=1] \mathbb P(D^*=1 \vert D=1)\\
	&\qquad \qquad + \mathbb E[Y_t(0)\vert D=1, D^*=0] \mathbb P(D^*=0 \vert D=1)\\
	&  \qquad \qquad - \mathbb E[Y_t(0)\vert D=0, D^*=1] \mathbb P(D^*=1 \vert D=0)\\
	&  \qquad \qquad - \mathbb E[Y_t(0)\vert D=0, D^*=0] \mathbb P(D^*=0 \vert D=0),\\
	&= \mathbb E[Y_t(0)\vert \varepsilon=0, D^*=1] \mathbb P(D^*=1 \vert D=1)\\
	&\qquad \qquad + \mathbb E[Y_t(0)\vert \varepsilon=1, D^*=0] \mathbb P(D^*=0 \vert D=1)\\
	&  \qquad \qquad - \mathbb E[Y_t(0)\vert \varepsilon=1, D^*=1] \mathbb P(D^*=1 \vert D=0)\\
	&  \qquad \qquad - \mathbb E[Y_t(0)\vert \varepsilon=0, D^*=0] \mathbb P(D^*=0 \vert D=0),
\end{eqnarray*}
where the first equality holds from the law of iterated expectations, and the second holds from the definition of the model. We can then take the difference of the above quantity over time as follows:
\begin{eqnarray*}
	&&\left\{\mathbb E[Y_1(0)\vert D=1]-\mathbb E[Y_1(0) \vert D=0]\right\}-\left\{\mathbb E[Y_0(0)\vert D=1]-\mathbb E[Y_0(0) \vert D=0]\right\}\\
	&=& \mathbb E[Y_1(0)-Y_0(0)\vert \varepsilon=0, D^*=1] \mathbb P(D^*=1 \vert D=1)\\
	&&\qquad \qquad + \mathbb E[Y_1(0)-Y_0(0)\vert \varepsilon=1, D^*=0] \mathbb P(D^*=0 \vert D=1)\\
	&&  \qquad \qquad - \mathbb E[Y_1(0)-Y_0(0)\vert \varepsilon=1, D^*=1] \mathbb P(D^*=1 \vert D=0)\\
	&&  \qquad \qquad - \mathbb E[Y_1(0)-Y_0(0)\vert \varepsilon=0, D^*=0] \mathbb P(D^*=0 \vert D=0)\\
	&=& \mathbb E[Y_1(0)-Y_0(0)\vert D^*=1] \mathbb P(D^*=1 \vert D=1)\\
	&&\qquad \qquad + \mathbb E[Y_1(0)-Y_0(0)\vert D^*=0] \mathbb P(D^*=0 \vert D=1)\\
	&&  \qquad \qquad - \mathbb E[Y_1(0)-Y_0(0)\vert D^*=1] \mathbb P(D^*=1 \vert D=0)\\
	&&  \qquad \qquad - \mathbb E[Y_1(0)-Y_0(0)\vert D^*=0] \mathbb P(D^*=0 \vert D=0),\\
	&=& \mathbb E[Y_1(0)-Y_0(0)\vert D^*=1] \left(\mathbb P(D^*=1 \vert D=1)-\mathbb P(D^*=1 \vert D=0)\right)\\
	&&\qquad \qquad + \mathbb E[Y_1(0)-Y_0(0)\vert D^*=0] \left(\mathbb P(D^*=0 \vert D=1)-\mathbb P(D^*=0 \vert D=0)\right),\\
	&=&\left\{\mathbb E[Y_1(0)-Y_0(0)\vert D^*=1]-\mathbb E[Y_1(0)-Y_0(0)\vert D^*=0]\right\} \left(\mathbb P(D^*=1 \vert D=1)-\mathbb P(D^*=1 \vert D=0)\right),\\
	&=& 0,
\end{eqnarray*}
where the second equality holds from Assumption \ref{PTepsDstar}.(\ref{PTeps}), the fourth follows from the fact that $\mathbb P(D^*=0 \vert D=1)-\mathbb P(D^*=0 \vert D=0)=-\left(\mathbb P(D^*=1 \vert D=1)-\mathbb P(D^*=1 \vert D=0)\right)$, and the last holds from Assumption \ref{PTepsDstar}.(\ref{PTDstar}).

\subsection{Proof of Proposition \ref{prop1}} We have
\begin{align*}
\theta^1_{OLS} &= \mathbb E[Y_1\vert D=1] - \mathbb E[Y_1 \vert D=0],\\
&= \mathbb E[Y_1(1) D^* +Y_1(0)(1-D^*) \vert D=1] -  \mathbb E[Y_1(1) D^* +Y_1(0)(1-D^*) \vert D=0],\\
&= \mathbb E[(Y_1(1)-Y_1(0)) D^* +Y_1(0) \vert D=1] -  \mathbb E[(Y_1(1)-Y_1(0)) D^* +Y_1(0) \vert D=0],\\
&= \mathbb E[(Y_1(1)-Y_1(0)) D^*\vert D=1] -  \mathbb E[(Y_1(1)-Y_1(0)) D^*\vert D=0]\\
& \qquad \qquad + \mathbb E[Y_1(0) \vert D=1]-\mathbb E[Y_1(0)\vert D=0],\\
&= \mathbb E[(Y_1(1)-Y_1(0)) \vert D^*=1, D=1] \mathbb P(D^*=1\vert D=1)\\
& \qquad \qquad -  \mathbb E[(Y_1(1)-Y_1(0))\vert D^*=1, D=0]\mathbb P(D^*=1\vert D=0)\\
& \qquad \qquad + \mathbb E[Y_1(0) \vert D=1]-\mathbb E[Y_1(0)\vert D=0],
\end{align*}
where the second equality holds from the definition of model (\ref{seq1}), and the last from the law of iterated expectations. Now, under Assumption \ref{PT} we have $$\mathbb E[Y_1(0) \vert D=1]-\mathbb E[Y_1(0) \vert D=0]=\mathbb E[Y_0\vert D=1] - \mathbb E[Y_0 \vert D=0].$$
Therefore,
\begin{align*}
\theta^1_{OLS} &= \mathbb E[(Y_1(1)-Y_1(0)) \vert D^*=1, D=1] \mathbb P(D^*=1\vert D=1)\\
& \qquad \qquad -  \mathbb E[(Y_1(1)-Y_1(0))\vert D^*=1, D=0]\mathbb P(D^*=1\vert D=0)\\
& \qquad \qquad + \theta^0_{OLS},
\end{align*}
which implies
\begin{align*}
\theta^1_{OLS}-\theta^0_{OLS} &= \mathbb E[(Y_1(1)-Y_1(0)) \vert D^*=1, D=1] \mathbb P(D^*=1\vert D=1)\\
& \qquad \qquad -  \mathbb E[(Y_1(1)-Y_1(0))\vert D^*=1, D=0]\mathbb P(D^*=1\vert D=0),
\end{align*}
that is,
\begin{align*}
\theta_{DID} &= \mathbb E[(Y_1(1)-Y_1(0)) \vert D^*=1, \varepsilon=0] \mathbb P(D^*=1\vert D=1)\\
& \qquad \qquad -  \mathbb E[(Y_1(1)-Y_1(0))\vert D^*=1, \varepsilon=1]\mathbb P(D^*=1\vert D=0),
\end{align*}
since $\{D^*=1, D=1\}=\{\varepsilon=0,D=1\}$ and $\{D^*=1, D=0\}=\{\varepsilon=1,D=0\}$ from the definition of model (\ref{seq1}).

\subsection{Proof of Corollary \ref{nofalseneg}}
From Proposition \ref{prop2}, we have
\begin{align*}
	ATT &= \frac{\mathbb P(D=1)}{\mathbb P(D^*=1)} \theta_{DID} + \frac{\mathbb E[(Y_1(1)-Y_1(0))\varepsilon \vert D^*=1]}{\mathbb P(D=0)}.
\end{align*}
When there are no false negatives, we have $\mathbb P(\varepsilon=1,D^*=1)=0$. Therefore,
\begin{align*}
	ATT &= \frac{\mathbb P(D=1)}{\mathbb P(D^*=1, \varepsilon=0)} \theta_{DID},\\
	&= \frac{\mathbb P(D=1)}{\mathbb P(D=1, \varepsilon=0)} \theta_{DID},
\end{align*}
which implies
\begin{align*}
	\theta_{DID} &= \frac{\mathbb P(D=1, \varepsilon=0)}{\mathbb P(D=1)} ATT=\mathbb P(\varepsilon=0\vert D=1) ATT.
\end{align*}

\subsection{Proof of equivalence between Assumption \ref{Mon} and the monotonicity condition in \cite{Hausman_al1998}}\label{apx:mon_equivalence}
We have
\begin{align*}
	\mathbb P(\varepsilon = 1 \vert D=1) &= \frac{\mathbb P(\varepsilon = 1, D=1)}{\mathbb P(D=1)}= \frac{\mathbb P(\varepsilon = 1, D^*=0)}{\mathbb P(D=1)},\\
	&= \frac{\mathbb P(\varepsilon = 1, D^*=0)}{\mathbb P(\varepsilon = 1, D=1)+\mathbb P(\varepsilon = 0, D=1)},\\
	&=  \frac{\mathbb P(\varepsilon = 1 \vert D^*=0) \mathbb P(D^*=0)}{\mathbb P(\varepsilon = 1, D^*=0)+\mathbb P(\varepsilon = 0, D^*=1)},
\end{align*}
where the first equality holds from Bayes' rule, the second holds from the definition of the model, and the third holds from the law of total probability and Bayes' rule. Similarly, we have
\begin{align*}
	\mathbb P(\varepsilon = 1 \vert D=1) &=  \frac{\mathbb P(\varepsilon = 1 \vert D^*=1) \mathbb P(D^*=1)}{\mathbb P(\varepsilon = 1, D^*=1)+\mathbb P(\varepsilon = 0, D^*=0)}.
\end{align*}
Denote $p=\mathbb P(D^*=1)$, $\alpha_0=\mathbb P(\varepsilon =1 \vert D^*=0)$, and $\alpha_1=\mathbb P(\varepsilon =1 \vert D=1)$.

Then,
\begin{align*}
	\mathbb P(\varepsilon = 1 \vert D=1) + \mathbb P(\varepsilon = 1 \vert D=0) < 1 &\Longleftrightarrow \frac{\alpha_0 (1-p)}{\alpha_0 (1-p) + (1-\alpha_1) p} + \frac{\alpha_1 p}{\alpha_1 p + (1-\alpha_0) (1-p)} < 1,\\
	&\Longleftrightarrow \alpha_0 (1-p)\alpha_1 p + \alpha_0 (1-p) (1-\alpha_0) (1-p)\\
	& + \alpha_1 p \alpha_0 (1-p) + \alpha_1 p (1-\alpha_1) p\\
	& < \alpha_0 (1-p) \alpha_1 p + \alpha_0 (1-p) (1-\alpha_0) (1-p)\\
	& + (1-\alpha_1) p \alpha_1 p + (1-\alpha_1) p (1-\alpha_0) (1-p),\\
	&\Longleftrightarrow \alpha_1 \alpha_0 < (1-\alpha_1) (1-\alpha_0),\\
	&\Longleftrightarrow \alpha_0 + \alpha_1 < 1.
\end{align*}

\subsection{Proof of Corollary \ref{attenuate}}
From Proposition \ref{prop1}, we have
\begin{align*}
\theta_{DID} &= \mathbb E\left[Y_1(1)-Y_1(0) \vert D^*=1, \varepsilon=0\right] \mathbb P(\varepsilon=0\vert D=1) \nonumber\\
& \qquad - \mathbb E\left[Y_1(1)-Y_1(0) \vert D^*=1, \varepsilon=1\right] \mathbb P(\varepsilon=1\vert D=0),\\
&= \mathbb E\left[Y_1(1)-Y_1(0) \vert D^*=1\right] \mathbb P(\varepsilon=0\vert D=1) \nonumber\\
& \qquad - \mathbb E\left[Y_1(1)-Y_1(0) \vert D^*=1\right] \mathbb P(\varepsilon=1\vert D=0),\\
&= \mathbb E\left[Y_1(1)-Y_1(0) \vert D^*=1\right]\left(\mathbb P(\varepsilon=0\vert D=1)-\mathbb P(\varepsilon=1\vert D=0)\right),\\
&= \mathbb E\left[Y_1(1)-Y_1(0) \vert D^*=1\right]\left(1-\mathbb P(\varepsilon=1\vert D=1)-\mathbb P(\varepsilon=1\vert D=0)\right),
\end{align*}
where the second equality holds under Assumption \ref{Nondiff}.

\subsection{Proof of Corollary \ref{bounds}}
From Corollary \ref{attenuate}, we have:
\begin{align*}
ATT &= \frac{\theta_{DID}}{1-\mathbb P(\varepsilon=1\vert D=1)-\mathbb P(\varepsilon=1\vert D=0)}.
\end{align*}
Therefore, the proof is straightforward from this equation and Assumption \ref{Upperbound}.

\subsection{Proof of Corollary \ref{fixedpoint}}
Suppose $\mathbb P(D^*=1)\neq \mathbb P(D=1)$. Then under Assumptions \ref{PT} and \ref{Nondiff}, we have from Propositions \ref{prop1} and \ref{prop2}:
\begin{align*}
\theta_{DID} &= \mathbb E\left[Y_1(1)-Y_1(0) \vert D^*=1, \varepsilon=0\right] \mathbb P(\varepsilon=0\vert D=1) \nonumber\\
& \qquad - \mathbb E\left[Y_1(1)-Y_1(0) \vert D^*=1, \varepsilon=1\right] \mathbb P(\varepsilon=1\vert D=0),\\
&= \mathbb E\left[Y_1(1)-Y_1(0) \vert D^*=1\right]\left(\mathbb P(\varepsilon=0\vert D=1)-\mathbb P(\varepsilon=1\vert D=0)\right)
\end{align*}
and
\begin{align*}
ATT &=\frac{\mathbb P(D=1)}{\mathbb P(D^*=1)} \theta_{DID} + \frac{\mathbb E[(Y_1(1)-Y_1(0))\varepsilon \vert D^*=1]}{\mathbb P(D=0)}.
\end{align*}
From the last equality, it follows that $ATT=\theta_{DID}$ if $\theta_{DID}=\frac{\mathbb E[(Y_1(1)-Y_1(0))\varepsilon \vert D^*=1]}{\mathbb P(D=0)\left(1-\frac{\mathbb P(D=1)}{\mathbb P(D^*=1)}\right)}$. On the other hand, $\mathbb E[(Y_1(1)-Y_1(0))\varepsilon \vert D^*=1]=\mathbb E[(Y_1(1)-Y_1(0))\vert D^*=1]\mathbb P(\varepsilon=1 \vert D^*=1)$. \\
Therefore, we have $ATT=\theta_{DID}$ if
\begin{align*}
 & \mathbb E\left[Y_1(1)-Y_1(0) \vert D^*=1\right]\left(1-\mathbb P(\varepsilon=1\vert D=1)-\mathbb P(\varepsilon=1\vert D=0)\right)\\
& = \frac{\mathbb E[(Y_1(1)-Y_1(0))\vert D^*=1]\mathbb P(\varepsilon=1 \vert D^*=1)}{\mathbb P(D=0)\left(1-\frac{\mathbb P(D=1)}{\mathbb P(D^*=1)}\right)}.
\end{align*}
A sufficient condition for the above equalities to hold is:
\begin{align*}
1-\mathbb P(\varepsilon=1\vert D=1)-\mathbb P(\varepsilon=1\vert D=0) = \frac{\mathbb P(\varepsilon=1 \vert D^*=1)}{\mathbb P(D=0)\left(1-\frac{\mathbb P(D=1)}{\mathbb P(D^*=1)}\right)},
\end{align*}
which is equivalent to
\begin{align*}
1-\mathbb P(\varepsilon=1\vert D=1)-\mathbb P(\varepsilon=1\vert D=0) = \frac{\mathbb P(\varepsilon=1, D^*=1)}{\mathbb P(D=0)\left(\mathbb P(D^*=1,\varepsilon=1)-\mathbb P(D^*=0, \varepsilon=1)\right)},
\end{align*}
since $$\mathbb P(D^*=1)=\mathbb P(D^*=1, D=1)+\mathbb P(D^*=1, \varepsilon=1),$$ and $$\mathbb P(D=1)=\mathbb P(D^*=1, D=1)+\mathbb P(D^*=0, \varepsilon=1).$$

\clearpage
\section{Combining Multiple Exogenous Data Sources for Identification}\label{combinedata}
\medskip

In this section, we are going to rely on the existence of additional data sources to study identification of some causal parameters of interest. Suppose that we have access to multiple data sets that are representative of the target population. Let $S$ denote a categorical variable indicator for each data source, and let $\mathcal S=\{0,1, \ldots, \bar{S}\}$ be the support of $S$.

\subsection{Identification}

We start with the mixture model representation. We have
\begin{eqnarray*}
\mathbb P(Y_t \leq y \vert D=1, S=s)&=&\mathbb P(Y_t \leq y, \varepsilon=1 \vert D=1, S=s) + \mathbb P(Y_t \leq y, \varepsilon=0 \vert D=1, S=s),\\
&=&\mathbb P(Y_t \leq y\vert  \varepsilon=1, D=1, S=s)  \mathbb P(\varepsilon=1\vert D=1, S=s) \\
&& \qquad \qquad +\  \mathbb P(Y_t \leq y\vert  \varepsilon=0, D=1, S=s)  \mathbb P(\varepsilon=0\vert D=1, S=s),\\
&=&\mathbb P(Y_t \leq y\vert  \varepsilon=1, D^*=0, S=s)  \mathbb P(\varepsilon=1\vert D=1, S=s) \\
&& \qquad \qquad +\  \mathbb P(Y_t \leq y\vert  \varepsilon=0, D^*=1, S=s)  \mathbb P(\varepsilon=0\vert D=1, S=s),\\
&=&\mathbb P(Y_t(0) \leq y\vert  \varepsilon=1, D^*=0, S=s)  \mathbb P(\varepsilon=1\vert D=1, S=s) \\
&& \qquad \qquad +\  \mathbb P(Y_t(1) \leq y\vert  \varepsilon=0, D^*=1, S=s)  \mathbb P(\varepsilon=0\vert D=1, S=s),\\
&=&\mathbb P(Y_t(0) \leq y\vert  \varepsilon=1, D^*=0)  \mathbb P(\varepsilon=1\vert D=1, S=s) \\
&& \qquad \qquad +\ \mathbb P(Y_t(1) \leq y\vert  \varepsilon=0, D^*=1)  \mathbb P(\varepsilon=0\vert D=1, S=s),
\end{eqnarray*}
where the first and second equalities hold from the law of total probability and Bayes' rule, respectively, the third follows from the definition of the model since $\{\varepsilon=1, D=1, S=s\}=\{\varepsilon=1, D^*=0, S=s\}$ and $\{\varepsilon=0, D=1, S=s\}=\{\varepsilon=0, D^*=1, S=s\}$, and the fourth and last hold under Assumption \ref{ass:datasourceexogeneity} and the definition of the potential outcome model.
Similarly, we can show the following result:
\begin{eqnarray*}
\mathbb P(Y_t \leq y \vert D=0, S=s)&=&\mathbb P(Y_t(1) \leq y\vert  \varepsilon=1, D^*=1)  \mathbb P(\varepsilon=1\vert D=0, S=s) \\
&& \qquad \qquad +\ \mathbb P(Y_t(0) \leq y\vert  \varepsilon=0, D^*=0)  \mathbb P(\varepsilon=0\vert D=0, S=s).
\end{eqnarray*}

We use the notation below throughout the rest of the paper.
\begin{notation}\label{notation}
Define $F_t(y\vert d,s) \equiv \mathbb P(Y_t \leq y \vert D=d, S=s)$, $F^{d^*}_t(y \vert \ell, d^*) \equiv \mathbb P(Y_t(d^*) \leq y\vert  \varepsilon=\ell, D^*=d^*)$, and $\alpha^d(s)\equiv \mathbb P(\varepsilon=1\vert D=d, S=s)$.
\end{notation}
From the above derivations, we have
\begin{eqnarray}
F_t(y\vert 1,s) &=& \alpha^1(s) F^0_t(y \vert 1,0) + (1-\alpha^1(s)) F^1_t(y \vert 0,1), \label{eq:mix1}\\
F_t(y\vert 0,s) &=& \alpha^0(s) F^1_t(y \vert 1,1) + (1-\alpha^0(s)) F^0_t(y \vert 0,0). \label{eq:mix2}
\end{eqnarray}
Our goal is to identify the distributions $F^0_t(y \vert 1,0),$ $F^1_t(y \vert 1,1),$ $F^1_t(y \vert 0,1),$ and $F^0_t(y \vert 0,0)$. Each equation defines a two-component mixture model where mixture weights vary with the data source $s$, while the mixture components do not.

Let us consider equation \eqref{eq:mix1} first. By differencing $F_t(y\vert 1,s)$ with respect to $s$, we have
\begin{eqnarray*}
F_t(y\vert 1,s_1^1) -F_t(y\vert 1,s_0^1) &=&  \left[\alpha^1(s_1^1)-\alpha^1(s_0^1)\right] \left[F^0_t(y \vert 1,0) - F^1_t(y \vert 0,1)\right].
\end{eqnarray*}
Under Assumption \ref{ass:relevance}, we can back out $F^0_t(y \vert 1,0)$ as a function of $F^1_t(y \vert 0,1)$:
\begin{eqnarray}
F^0_t(y \vert 1,0) &=& F^1_t(y \vert 0,1) + \frac{1}{\left[\alpha^1(s_1^1)-\alpha^1(s_0^1)\right]} \left[ F_t(y\vert 1,s_1^1) -F_t(y\vert 1,s_0^1) \right].  \label{eq:mix1ident1}
\end{eqnarray}
Plugging \eqref{eq:mix1ident1} in \eqref{eq:mix1} for $s=s_0^1$ yields
\begin{eqnarray*}
F^1_t(y \vert 0,1) &=& F_t(y \vert 1,s_0^1) - \frac{\alpha^1(s_0^1)}{\left[\alpha^1(s_1^1)-\alpha^1(s_0^1)\right]} \left[ F_t(y\vert 1,s_1^1) -F_t(y\vert 1,s_0^1) \right].
\end{eqnarray*}
Therefore, the distributions $F^1_t(y \vert 0,1)$ and $F^0_t(y \vert 1,0)$ are identified up to a scalar parameter:
\begin{eqnarray}
F^1_t(y \vert 0,1) &=& F_t(y \vert 1,s_0^1) -\eta^1 \left[ F_t(y\vert 1,s_1^1) -F_t(y\vert 1,s_0^1) \right],\label{eq:mix:id1}\\
F^0_t(y \vert 1,0) &=& F_t(y \vert 1,s_0^1) + (\theta^1-\eta^1) \left[ F_t(y\vert 1,s_1^1) -F_t(y\vert 1,s_0^1) \right], \label{eq:mix:id2}
\end{eqnarray}
where $\eta^1\equiv \frac{\alpha^1(s_0^1)}{\left[\alpha^1(s_1^1)-\alpha^1(s_0^1)\right]}$ and $\theta^1\equiv \frac{1}{\left[\alpha^1(s_1^1)-\alpha^1(s_0^1)\right]}$. At this point, if we can identify $\eta^1$ and $\theta^1$, we are done. Let us denote $\Lambda^1(s)\equiv \frac{\alpha^1(s)-\alpha^1(s_0^1)}{\alpha^1(s_1^1)-\alpha^1(s_0^1)}$. Then, the weight function $\alpha^1(s)$ can be written as
\begin{eqnarray}\label{eq:mix:id3}
\alpha^1(s)=\frac{1}{\theta^1}\left(\eta^1+\Lambda^1(s)\right).
 \end{eqnarray}
 If $F^0_t(y \vert 1,0)=F^1_t(y \vert 0,1)$ for all $y$, then they are identified as $F_t(y\vert 1,s)$, which must not vary with $s$, i.e., $F_t(y\vert 1,s)=F_t(y\vert 1)$. If there exists $y^1$ such that $F_t(y^1\vert 1,s_1^1)\neq F_t(y^1\vert 1,s_0^1)$,  then $\Lambda^1(s)$ is identified as
$$\Lambda^1(s)=\frac{F_t(y^1\vert 1,s) - F_t(y^1\vert 1,s_0^1)}{F_t(y^1\vert 1,s_1^1) - F_t(y^1\vert 1,s_0^1)}.$$
Finally, we partially-identify $\eta^1$ and $\theta^1$ using the constraints that $F^1_t(y \vert 0,1)$ and $F^0_t(y \vert 1,0)$ are cumulative distribution functions (cdfs), and $\alpha^1(s)$ is a probability weight. It is clear from \eqref{eq:mix:id1} and \eqref{eq:mix:id2} that these functions are right-continuous, have their limits at $-\infty$ and $\infty$ equal to 0 and 1, respectively. It remains to check the monotonicity condition. The following theorem holds.

\begin{theorem}\label{thm:id1}
Under Assumptions \ref{ass:datasourceexogeneity}-\ref{ass:relevance}, the distributions $F^1_t(y \vert 0,1)$, $F^0_t(y \vert 1,0)$, and $\alpha^1(s)$ are identified as in Equations \eqref{eq:mix:id1}, \eqref{eq:mix:id2}, and \eqref{eq:mix:id3}, respectively, where $\eta^1$ and $\theta^1$ are partially-identified from the constraints below: for all $(y, s)$,
\begin{eqnarray}
&& f_t(y \vert 1,s_0^1) -\eta^1 \left[ f_t(y\vert 1,s_1^1) -f_t(y\vert 1,s_0^1) \right]\geq 0,\label{eq:mix:id4}\\
&& f_t(y \vert 1,s_0^1) + (\theta^1-\eta^1) \left[ f_t(y\vert 1,s_1^1) -f_t(y\vert 1,s_0^1) \right] \geq 0, \label{eq:mix:id5}\\
&&0\leq \frac{1}{\theta^1}\left(\eta^1+\Lambda^1(s)\right) \leq 1,\label{eq:mix:id6}
\end{eqnarray}
where $f_t(y\vert d,s)$ denotes the density or probability mass function of $Y_t$ conditional on $(D=d,S=s)$, depending on whether $Y_t$ is continuous, discrete or mixed.

Similar identification results hold for $F^0_t(y \vert 0,0)$, $F^1_t(y \vert 1,1)$, and $\alpha^0(s)$.
\end{theorem}
Now that we partially-identify $F^1_t(y \vert 0,1)$, $F^0_t(y \vert 1,0)$,  $F^0_t(y \vert 0,0)$, and $F^1_t(y \vert 1,1)$, we can subsequently partially-identify $\mathbb E[Y_t(1) \vert \varepsilon=0,D^*=1]$, $\mathbb E[Y_t(0) \vert \varepsilon=1,D^*=0]$,  $\mathbb E[Y_t(0) \vert \varepsilon=0,D^*=0]$, and $\mathbb E[Y_t(1) \vert \varepsilon=1,D^*=1]$. From there, we can partially-identify the ATT for the correctly classified and the misclassified subpopulations under the parallel trends and no-anticipation effects assumptions.

\begin{proposition}\label{prop:idATTepsilons}
Under Assumptions \ref{ass:datasourceexogeneity}-\ref{PTdstarallgroups}, $ATT_{\varepsilon=\ell}\equiv \mathbb E[Y_1(1)-Y_1(0) \vert \varepsilon=\ell, D^*=1]$ is partially-identified as
$$\left(\mathbb E[Y_1\vert \varepsilon=\ell, D^*=1]-\mathbb E[Y_1\vert \varepsilon=\ell, D^*=0]\right)-\left(\mathbb E[Y_0\vert \varepsilon=\ell, D^*=1]-\mathbb E[Y_0\vert \varepsilon=\ell, D^*=0]\right),$$
where the distributions $F^1_t(y \vert 0,1)$, $F^0_t(y \vert 1,0)$,  $F^0_t(y \vert 0,0)$, and $F^1_t(y \vert 1,1)$ are partially-identified from Theorem \ref{thm:id1}.
\end{proposition}

\subsection{Estimation}
In order to estimate the identified sets for $\eta^1$ and $\theta^1$, we convert inequalities \eqref{eq:mix:id4}-\eqref{eq:mix:id6} into conditional moment inequalities, which allow us to implement the method using existing papers such as \cite{CLR2013} and \cite{CKLRstata}. Let us consider inequality \eqref{eq:mix:id4} first. From Bayes' rule, we have $$f_t(y \vert d,s)=\frac{f_t(y)*\mathbb P(D=d,S=s\vert Y_t=y)}{\mathbb P(D=d,S=s)}.$$
Therefore, inequality \eqref{eq:mix:id4} is equivalent to:
\begin{eqnarray*}
&&\frac{f_t(y) \mathbb P(D=1,S=s^1_0\vert Y_t=y)}{\mathbb P(D=1,S=s^1_0)} \\
&& \qquad \qquad -\eta^1 \left[\frac{f_t(y) \mathbb P(D=1,S=s^1_1\vert Y_t=y)}{\mathbb P(D=1,S=s^1_1)} -\frac{f_t(y) \mathbb P(D=1,S=s^1_0\vert Y_t=y)}{\mathbb P(D=1,S=s^1_0)} \right] \geq 0.
\end{eqnarray*}
When $f_t(y)=0$, the inequality holds trivially with equality and does not bring any identifying power to the identified set for $\eta^1$. So, we focus on the case where $f_t(y)>0$. Hence, after dropping $f_t(y)$ from all sides, inequality \eqref{eq:mix:id4} becomes
\begin{eqnarray*}
&& \frac{\mathbb P(D=1,S=s^1_0\vert Y_t=y)}{\mathbb P(D=1,S=s^1_0)} \\
&& \qquad \qquad -\eta^1 \left[\frac{\mathbb P(D=1,S=s^1_1\vert Y_t=y)}{\mathbb P(D=1,S=s^1_1)} -\frac{\mathbb P(D=1,S=s^1_0\vert Y_t=y)}{\mathbb P(D=1,S=s^1_0)} \right] \geq 0,
\end{eqnarray*}
which in turn is equivalent to
\begin{eqnarray*}
\inf_{y\in \mathcal Y}\mathbb E \bigg[ \frac{D\mathbbm{1}\{S=s^1_0\}}{\mathbb E[D\mathbbm{1}\{S=s^1_0\}]} -\eta^1 \left(\frac{D\mathbbm{1}\{S=s^1_1\}}{\mathbb E[D\mathbbm{1}\{S=s^1_1\}]} -\frac{D\mathbbm{1}\{S=s^1_0\}}{\mathbb E[D\mathbbm{1}\{S=s^1_0\}]} \right) \vert Y_t=y \bigg] \geq 0.
\end{eqnarray*}

Similarly, inequality \eqref{eq:mix:id5} can equivalently be rewritten as
\begin{eqnarray*}
\inf_{y\in \mathcal Y}\mathbb E \bigg[ \frac{D\mathbbm{1}\{S=s^1_0\}}{\mathbb E[D\mathbbm{1}\{S=s^1_0\}]} + (\theta^1-\eta^1) \left(\frac{D\mathbbm{1}\{S=s^1_1\}}{\mathbb E[D\mathbbm{1}\{S=s^1_1\}]} -\frac{D\mathbbm{1}\{S=s^1_0\}}{\mathbb E[D\mathbbm{1}\{S=s^1_0\}]} \right) \vert Y_t=y \bigg] \geq 0.
\end{eqnarray*}

Now, we are going to transform inequality \eqref{eq:mix:id6} into two conditional moment inequalities. Define $\text{sign}(\theta^1)\equiv \mathbbm{1}\{\theta^1 > 0\}-\mathbbm{1}\{\theta^1 < 0\}$. Since $\theta^1 \neq 0$, we have $\theta^1 \text{sign}(\theta^1) > 0$. By multiplying each side of \eqref{eq:mix:id6} by $\theta^1 \text{sign}(\theta^1)$, we obtain:
\begin{eqnarray*}
&& \text{sign}(\theta^1)(\eta^1+\Lambda^1(s)) \geq 0,\\
&& \text{sign}(\theta^1)(\theta^1-\eta^1-\Lambda^1(s)) \geq 0,
\end{eqnarray*}
which we equivalently rewrite as
\begin{eqnarray*}
&&\inf_{s\in \mathcal S}\mathbb E\bigg[ \text{sign}(\theta^1)\bigg(\eta^1\nonumber\\
&& \qquad +\frac{\mathbbm{1}\{Y_t \leq y^1\}-\mathbb E[\mathbbm{1}\{Y_t \leq y^1\}\vert D=1,S=s^1_0]}{\mathbb E[\mathbbm{1}\{Y_t \leq y^1\} \vert D=1,S=s^1_1]-\mathbb E[\mathbbm{1}\{Y_t \leq y^1\}\vert D=1,S=s^1_0]}\bigg) \vert D=1, S=s\bigg ] \geq 0,\\
&& \inf_{s\in \mathcal S} \mathbb E\bigg[ \text{sign}(\theta^1)\bigg(\theta^1-\eta^1 \nonumber\\
&& \qquad -\frac{\mathbbm{1}\{Y_t \leq y^1\}-\mathbb E[\mathbbm{1}\{Y_t \leq y^1\}\vert D=1,S=s^1_0]}{\mathbb E[\mathbbm{1}\{Y_t \leq y^1\} \vert D=1,S=s^1_1]-\mathbb E[\mathbbm{1}\{Y_t \leq y^1\}\vert D=1,S=s^1_0]}\bigg) \vert D=1, S=s\bigg ] \geq 0.
\end{eqnarray*}

\subsection{Proof of Proposition \ref{prop:ATT_uncond}:}

\begin{proof}
		For \eqref{eq:ATT_uncond_decomp}, apply the law of iterated expectations conditional on $\varepsilon$:
		\[
		ATT
		=
		\E\!\left[Y_1(1)-Y_1(0)\mid D^*=1\right]
		=
		\sum_{\ell\in\{0,1\}}\E\!\left[Y_1(1)-Y_1(0)\mid \varepsilon=\ell,D^*=1\right]\Pp(\varepsilon=\ell\mid D^*=1),
		\]
		which equals $(1-p)ATT_{\varepsilon=0}+pATT_{\varepsilon=1}$.

		For \eqref{eq:p_formula}, note that $\{\varepsilon=1,D^*=1\}$ implies $\{\varepsilon=1,D=0\}$ by \eqref{seq1}, hence
		\begin{eqnarray*}
		\Pp(\varepsilon=1,D^*=1)&=&\sum_{s\in\mathcal S}\Pp(D=0,\varepsilon=1\mid S=s)\Pp(S=s)\\
		&=&\sum_{s\in\mathcal S}\Pp(\varepsilon=1\mid D=0,S=s)\Pp(D=0\mid S=s)\Pp(S=s).
		\end{eqnarray*}
		But $\Pp(\varepsilon=1\mid D=0,S=s)=\alpha^0(s)$ and $\Pp(D=0\mid S=s)=1-\pi(s)$, so the numerator in \eqref{eq:p_formula} follows.

		Similarly,
		\[
		\Pp(D^*=1)
		=
		\sum_{s\in\mathcal S}\Pp(D^*=1\mid S=s)\Pp(S=s),
		\]
		and by the law of total probability over $D$,
		\begin{eqnarray*}
		\Pp(D^*=1\mid S=s)&=&\Pp(D^*=1\mid D=1,S=s)\Pp(D=1\mid S=s)\\
        &&+\Pp(D^*=1\mid D=0,S=s)\Pp(D=0\mid S=s).
		\end{eqnarray*}
		Using $\Pp(D^*=1\mid D=1,S=s)=\Pp(\varepsilon=0\mid D=1,S=s)=1-\alpha^1(s)$ and $\Pp(D^*=1\mid D=0,S=s)=\Pp(\varepsilon=1\mid D=0,S=s)=\alpha^0(s)$ gives the denominator in \eqref{eq:p_formula}.
		Finally, $\mathcal A$ is the collection of all convex combinations \eqref{eq:ATT_uncond_decomp} over $a_0\in\mathcal A_0$, $a_1\in\mathcal A_1$,
		and $p\in\mathcal P$.
	\end{proof}

    \vspace{0.5cm}
\subsection{Proof of Proposition \ref{prop:ATT_uncond_prime}:}

\begin{proof}
			By definition, for each $\ell\in\{0,1\}$,
			\[
			ATT_{\varepsilon=\ell}
			=
			\E\!\left[Y_1(1)-Y_1(0)\mid \varepsilon=\ell,D^*=1\right].
			\]
			Under Assumption \ref{Nondiff}, $\varepsilon\indep (Y_1(1),Y_1(0))\mid D^*$, so in particular conditioning on $D^*=1$ implies
			\[
			\E\!\left[Y_1(1)-Y_1(0)\mid \varepsilon=\ell,D^*=1\right]
			=
			\E\!\left[Y_1(1)-Y_1(0)\mid D^*=1\right]
			=
			ATT,
			\]
			for both $\ell=0$ and $\ell=1$. Hence $ATT\in\mathcal A_0$ and $ATT\in\mathcal A_1$, so $ATT\in\mathcal A_0\cap\mathcal A_1\equiv \mathcal A^{ND}$.
			The set inclusion $\mathcal A^{ND}\subseteq \mathcal A$ follows because Proposition \ref{prop:ATT_uncond} allows $(1-p)a_0+pa_1$ to vary over
			$a_0\in\mathcal A_0$, $a_1\in\mathcal A_1$, and $p\in\mathcal P$, whereas under Assumption \ref{Nondiff} the only admissible values satisfy $a_0=a_1=ATT$.
		\end{proof}

\clearpage
\section{Simulation Results for Nondifferential Misclassification}\label{sim-nondifferential}

\begin{table}[htp!]
	\centering
	\caption{\textbf{Simulation Results for Nondifferential Misclassification of Treatment}}
	\label{tab2}
	\begin{threeparttable}
        \footnotesize
\begin{tabular}{@{}cccccc@{}}
\toprule
\multirow[c]{2}{*}{\textbf{Overall error rate}} & \multirow[c]{2}{*}{\parbox{2cm}{\centering \textbf{False negative}}} & \multirow[c]{2}{*}{\parbox{2cm}{\centering \textbf{False positive}}} & \multicolumn{2}{c}{\textbf{DID estimates}} & \multirow[c]{2}{*}{\parbox{2.5cm}{\centering \textbf{ATT bounds ($\lambda=0.4$)}}}  \\
\cmidrule{4-5}      &       &       & \textbf{True treatment} & \textbf{Observed treatment} & \\
\addlinespace
& (1) &(2) & (3) &  (4) & (5) \\
\midrule
\multicolumn{3}{l}{\underline{Panel A: Symmetric errors}} &       &       &         \\
5\%   & 0.050 & 0.050 & 2.999 & 2.700 & (2.700 , 4.499) \\
10\%   & 0.100 & 0.100 & 3.000 & 2.400 & (2.400 , 4.000) \\
20\%   & 0.200 & 0.200 & 3.000 & 1.800 & (1.800 , 3.001) \\
30\%   & 0.300 & 0.300 & 3.000 & 1.200 & (1.200 , 2.001) \\
40\%   & 0.400 & 0.400 & 3.000 & 0.600 & (0.600 , 1.000) \\
50\%   & 0.490 & 0.490 & 3.001 & 0.060 & (0.060 , 0.100) \\
\hline 
\addlinespace
\multicolumn{3}{l}{\underline{Panel B: Asymmetric errors}} &       &       &         \\
5\%   & 0.011 & 0.069 & 3.000 & 2.760 & (2.760 , 4.599) \\
10\%   & 0.170 & 0.028 & 3.000 & 2.408 & (2.408 , 4.013) \\
20\%   & 0.288 & 0.019 & 2.999 & 2.077 & (2.077 , 3.461) \\
30\%   & 0.376 & 0.015 & 3.000 & 1.826 & (1.826 , 3.044) \\
40\%   & 0.445 & 0.012 & 3.000 & 1.629 & (1.629 , 2.715) \\
50\%   & 0.487 & 0.010 & 3.000 & 1.511 & (1.511 , 2.518) \\
\hline 
\addlinespace
\multicolumn{3}{l}{\underline{Panel C: False negatives only}} &       &       &         \\
5\%   & 0.091 & 0.000 & 3.000 & 2.727 & (2.727 , 4.545) \\
10\%   & 0.167 & 0.000 & 3.000 & 2.501 & (2.501 , 4.168) \\
20\%   & 0.286 & 0.000 & 3.000 & 2.143 & (2.143 , 3.572) \\
30\%   & 0.375 & 0.000 & 2.999 & 1.875 & (1.875 , 3.124) \\
40\%   & 0.445 & 0.000 & 3.000 & 1.667 & (1.667 , 2.778) \\
50\%   & 0.487 & 0.000 & 3.000 & 1.535 & (1.535 , 2.558) \\
\hline 
\addlinespace
\multicolumn{3}{l}{\underline{Panel D: False positives only}} &       &       &         \\
5\%   & 0.000 & 0.091 & 3.001 & 2.728 & (2.728 , 4.547) \\
10\%   & 0.000 & 0.167 & 3.000 & 2.500 & (2.500 , 4.167) \\
20\%   & 0.000 & 0.286 & 3.000 & 2.143 & (2.143 , 3.571) \\
30\%   & 0.000 & 0.375 & 3.000 & 1.875 & (1.875 , 3.124) \\
40\%   & 0.000 & 0.444 & 3.000 & 1.667 & (1.667 , 2.778) \\
50\%   & 0.000 & 0.487 & 3.000 & 1.538 & (1.538 , 2.563) \\
\bottomrule 
\end{tabular}

		\begin{tablenotes}[flushleft]
			\footnotesize
            \setlength{\baselineskip}{11pt}
			\item \textbf{Notes:} This table presents simulation results for the case of nondifferential misclassification of treatment within the difference-in-differences framework. Columns 1 and 2 report false negative and false positive rates conditional on true treatment status. The DID estimates using the true (unobserved) and misclassified (observed) treatment variables are reported in Columns 3 and 4, respectively. The bounds on the ATT in Column 5 are point-estimate bounds based on Corollary \ref{bounds} in the main paper. The panels correspond to various types of misclassification discussed in the Monte Carlo design setup section of the main paper.
		\end{tablenotes}
	\end{threeparttable}
\end{table}

\clearpage
\section{Empirical Illustration: Punishment Severity and Jury Decisions}\label{jury-appendix}

\citet{bindler2018punishment} use two natural experiments in English history---the abolition of capital punishment and the temporary halt of transportation during the American Revolution---to investigate the effect of a large change in punishment severity on jury decision making. We use the capital punishment experiment in our empirical illustration because those data contain treatment and control groups permitting a DID analysis. The basic idea is to exploit a series of offense-specific laws in the middle of the 19\textsuperscript{th} century that abolished capital punishment in England to study the impact of sharp reduction in punishment severity on a jury's sentencing decisions. The data are extracted from more than 200,000 criminal cases tried at the Old Bailey Criminal Court in London between 1772 and 1871. In addition to sentencing outcomes (verdicts), the data includes basic information on each case including the alleged offense, the demographic information on the defendant (name, age, and gender), the session dates, and the names of the judges and juries.

To implement the DID research design, \citet{bindler2018punishment} compare changes in sentencing outcomes for cases in which capital punishment was abolished for the alleged offense (``treated'' offenses) to those cases where the alleged offense was never eligible for capital punishment or always capital eligible (``control'' offenses). The authors find that the drastic reduction in punishment severity following the abolishing of capital punishment led to a significant increase in the chance of any conviction and conviction of the original charge. Their results also indicate a decrease in the recommendation for a mercy following the abolishing of capital punishment since this was no longer necessary for juries wanting to spare a defendant of death, especially in violent and fraud cases.

A key empirical challenge in the original analysis is the inability to unambiguously identify treated offense categories and the timing of the treatment years because of the complicated nature of the historical laws spanning almost 200 years. The authors resort to using the observed discontinuities in the share of death sentences to identify treated and control offenses. Specifically, an offense is coded as a ``treated'' offense when the share of death sentences drops to zero, with the treatment period starting from the year this discontinuity was observed. Otherwise, if no such discontinuity occurs (i.e., always or never capital eligible), then the offense was assigned to the ``control'' group. Thus, misclassification may arise due to the lag between the passage and implementation of the offense-specific law abolishing capital punishment, especially since the authors cannot code treatment based on the exact month of the law. Admittedly, the authors note that ``our coding of the reform as the first year with zero death sentences therefore likely assigns some `treated' cases to the pre-reform period'' but consider this of minimal concern given their robustness checks \citep{bindler2018punishment}. Their final treatment and control groups comprised sixteen and nine offense categories, respectively.

\begin{figure}[htp!]
	\caption{Identifying the timing of the abolition of capital punishment (treated offenses)} \label{treatment-timing}
	\begin{subfigure}{0.49\textwidth}
		\centering
		\caption{Animal theft}
		\includegraphics[width=0.8\linewidth]{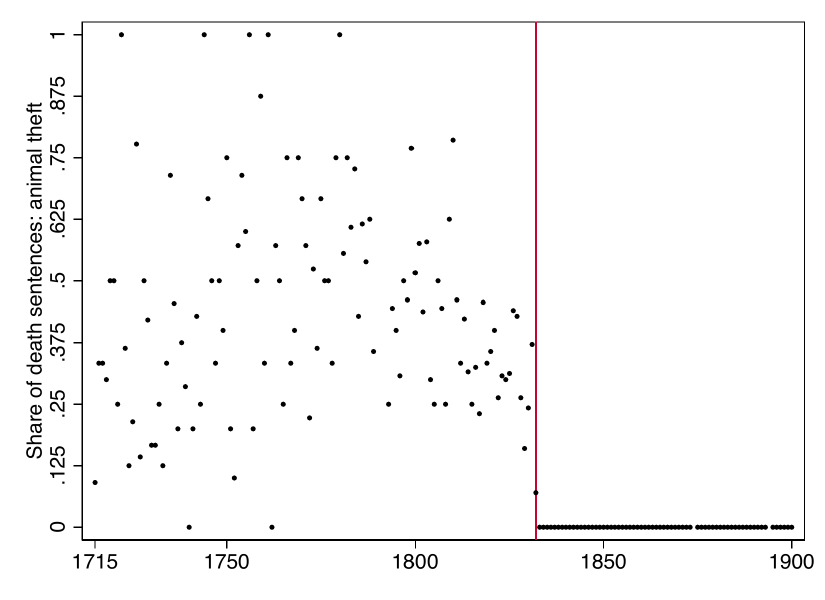}
		\label{fig:animal-theft}
	\end{subfigure}
	\begin{subfigure}{0.49\textwidth}
		\centering
		\caption{Coining offenses}
		\includegraphics[width=0.8\linewidth]{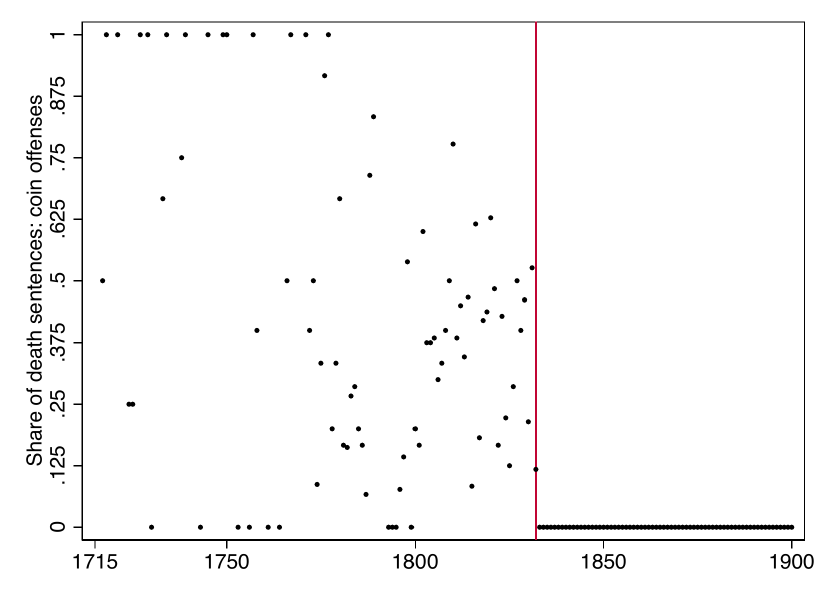}
		\label{fig:coining-offenses}
	\end{subfigure}
	\newline
	\begin{subfigure}{0.49\textwidth}
		\centering
		\caption{Forgery}
		\includegraphics[width=0.8\linewidth]{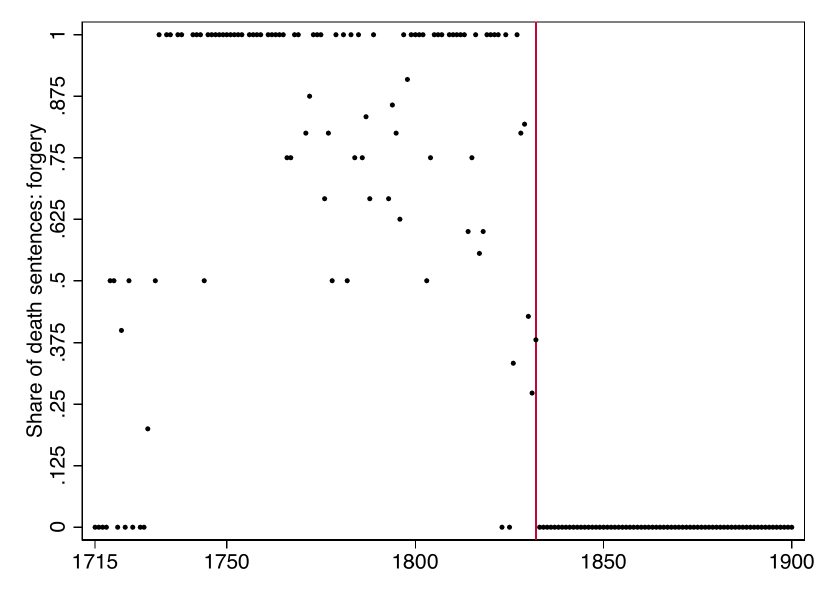}
		\label{fig:forgery}
	\end{subfigure}
	\begin{subfigure}{0.49\textwidth}
		\centering
		\caption{Sodomy}
		\includegraphics[width=0.8\linewidth]{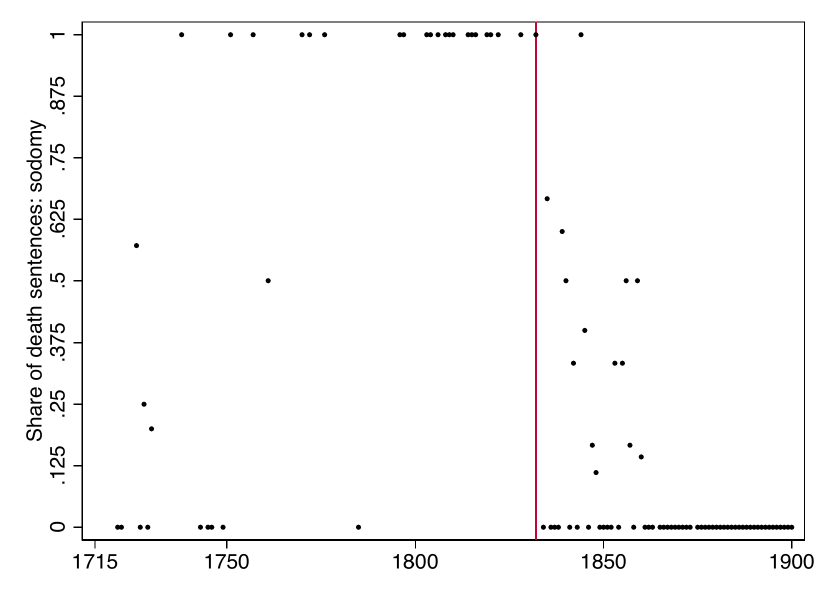}
		\label{fig:sodomy}
	\end{subfigure}
	\newline

	\begin{subfigure}{0.49\textwidth}
		\centering
		\caption{Theft from place}
		\includegraphics[width=0.8\linewidth]{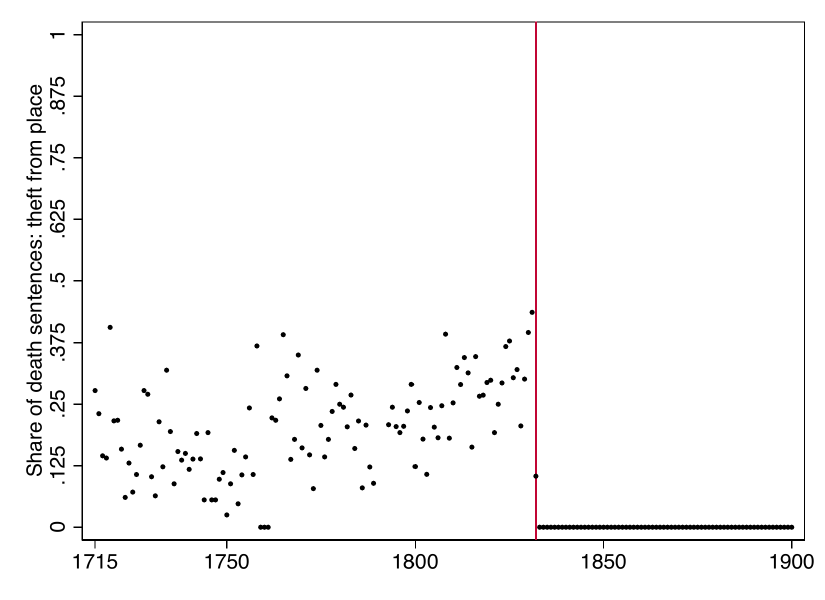}
		\label{fig:theft-from-place}
	\end{subfigure}

	\vspace{-0.5cm}
	\begin{tablenotes}[flushleft]
		\footnotesize
		\begin{singlespace}
			\item \textbf{Notes:}  The figure displays the proportion of convicted cases that resulted in a death sentence based on data  \cite{bindler2018punishment}. The treated offenses are animal theft (Panel a), coining offenses (Panel b), forgery (Panel c), sodomy (Panel d), and theft from place (Panel e). The red vertical line denotes the year of treatment (1832), which is the first year in which the capital sentences drop to zero.
		\end{singlespace}
	\end{tablenotes}
\end{figure}

In our replication, we use the same control group offenses but only include five offense categories in our treatment group. We do so because capital punishment was abolished in different years for many of the treated offense categories over the study period, leading to a staggered treatment adoption DID setup that is outside the scope of this paper. However, the authors determined that the abolishing of capital punishment occurred in the same year (1832) for the five offense categories we use to define treatment. The five offense categories are animal theft, coining offenses, forgery, sodomy, and theft from place. Figure \ref{treatment-timing} displays the share of death sentences for our treated offenses over time with the solid (red) line denoting the year capital punishment was abolished. All the offenses appear to have witnessed a sharp discontinuous drop in the share of death sentences by 1832 although there are some non-zero death sentences for sodomy post-1832.

Table \ref{tab4} presents our results for the impact of abolishing capital punishment on jury decision making. Similar to our first application, we perform all analyses without covariates and maintain the parallel trends assumption based on the justification and evidence provided in the original study. The table reports the standard DID estimates (with no accounting for misclassification) and two sets of our proposed bounds accounting for potential misclassification of treatment based on two estimates of $\lambda$. The first set of bounds (Column 2) estimate $\lambda$ as the proportion of cases in the years of potential misclassification---pre-reform (1831), reform (1832), and post-reform (1833)---to the number of cases in the entire study period 1803-1871, leading to $\lambda=0.07$. The second set of bounds (Column 3) estimates $\lambda$ using the same numerator as before, but dividing by the number of cases in the post-intervention period spanning 1831-1871; this yields $\lambda=0.125$.

\begin{table}[htp!]
	\centering
	\caption{\textbf{The Impact of Abolishing Capital Punishment on Jury Decisions}}
	\label{tab4}
	\begin{threeparttable}
		\begin{tabular}{@{}p{5cm}ccc@{}}
\toprule
\multicolumn{1}{c}{\multirow{2}[4]{*}{\textbf{Dependent variables}}} & \multicolumn{1}{c}{\multirow{2}[4]{*}{\textbf{DID estimates}}} & \multicolumn{2}{P{4cm}}{\textbf{ATT bounds}} \\
\cmidrule{3-4}      &       & \multicolumn{1}{P{2cm}}{$\lambda = 0.07$} & \multicolumn{1}{P{2cm}}{$\lambda = 0.125$} \\
\addlinespace
& (1) &(2) & (3)  \\
\midrule
\emph{Panel A: Full sample}&  & &  \\
\addlinespace
Guilty of any offense by jury verdict (0/1) & 0.0655  &  (0.0655 , 0.0705)  &  (0.0655 , 0.0749)  \\
& (0.0670) &       &  \\
\addlinespace
Guilty of original charge by jury verdict (0/1) & 0.1656***  &  (0.1656 , 0.1781)  &  (0.1656 , 0.1893)  \\
& (0.0406) &       &  \\
\addlinespace
\emph{Panel B: Sample of convicted cases}&  & &  \\
\addlinespace
Guilty of lesser offense conditional on guilty by jury verdict (0/1), broad definition & -0.1537*  &  (-0.1653 , -0.1537)  &  (-0.1757 , -0.1537)  \\
 & (0.0855) &       &  \\
\addlinespace
Recommended for mercy conditional on guilty by jury verdict (0/1) & -0.0495**  &  (-0.0532 , -0.0495)  &  (-0.0566 , -0.0495)  \\
 & (0.0166) &       &  \\
\addlinespace
\bottomrule 
\end{tabular}%

		\begin{tablenotes}[flushleft]
			\footnotesize
            \setlength{\baselineskip}{11pt}
			\item \textbf{Notes:} This table presents difference-in-differences estimates of the effect of changes in punishment severity due to abolishing offense-specific capital punishment on jury decision-making \cite{bindler2018punishment}. The sample comprises defendant-case observations tried in the Old Bailey Criminal Court in London between 1772 and 1871. All analyses are conducted without covariates. Column 1 reports the baseline DID estimates. The bounds on the ATT in Columns (2) and (3) are point-estimate bounds based on Corollary \ref{bounds} of the main paper that accounts for misclassification. The full sample (Panel A) and the sample of convicted cases comprise 80,925 and 60,214 observations, respectively.
		\end{tablenotes}
	\end{threeparttable}
\end{table}

Although we define our treatment group differently to include only offenses treated in the same year of 1832, the DID estimates in Column 1 are qualitatively the same as those in the original paper (using all the treated offenses) except in one instance. The DID estimates show that abolishing capital punishment led to a 6.5 percentage increase in the chance of conviction of any charge (original or lesser charge) but this is not statistically significant.\footnote{Since \cite{bindler2018punishment} find that their overall results for the probability of any charge are driven by violent offenses, the fact that we do not find a statistically significant effect is not surprising given our definition of treatment includes mostly non-violent offenses.} However, we find a statistically significant increase in the probability of any conviction of the original charge by 16.5 percentage points. For the next two outcomes (Panel B), the sample is restricted to those who were convicted of their alleged offenses (as in the original study). We find that the abolishing of capital punishment reduces the probability of lesser offense conditional on being convicted by 15.4 percentage points. Finally, abolishing capital punishment reduces the chance of recommendation for mercy conditional on being convicted by 4.9 percentage points.

Our two sets of bounds show how sensitive the DID estimates are to the potential misclassification of treatment. The results in Column 2 show that the DID estimates are biased downwards by about 7 percent in all cases. For a higher rate of misclassification in Column 3, our bounds are slightly wider, with an implied downward bias of about 12.5 percent. Taken together, our proposed bounds can help researchers gauge the robustness of their findings to a potentially misclassified treatment in DID settings with no access to the true treatment.

    \medskip

\end{document}